\documentclass[lettersize,journal]{IEEEtran}
\usepackage{amsmath,amsfonts}
\usepackage{algorithmic}
\usepackage{array}
\usepackage[caption=false,font=normalsize,labelfont=sf,textfont=sf]{subfig}
\usepackage{textcomp}
\usepackage{stfloats}
\usepackage{url}
\usepackage{verbatim}
\usepackage{graphicx}
\usepackage{arydshln}
\usepackage{makecell}
\usepackage{hyperref}
\usepackage{subfig}
\usepackage{threeparttable}
\usepackage{multicol}
\usepackage{multirow}
\usepackage{adjustbox}
\usepackage[capitalize]{cleveref}
\def\BibTeX{{\rm B\kern-.05em{\sc i\kern-.025em b}\kern-.08em
    T\kern-.1667em\lower.7ex\hbox{E}\kern-.125emX}}
\usepackage{balance}

\begin{document}
\title{Uni-LVC: A Unified Method for Intra- and Inter-Mode\\ Learned Video Compression}
\author{
    Yichi Zhang,~Ruoyu Yang,~and Fengqing Zhu%
    \thanks{
    Yichi Zhang, Ruoyu Yang, and Fengqing Zhu are with the Elmore Family School of Electrical and Computer Engineering, Purdue University, West Lafayette, IN 47907 USA. (e-mail: zhan5096@purdue.edu; yang2729@purdue.edu; zhu0@purdue.edu) 
    }%
}

\markboth{Journal of \LaTeX\ Class Files,~Vol.~18, No.~9, September~2020}%
{How to Use the IEEEtran \LaTeX \ Templates}

\maketitle

\begin{abstract}
Recent advances in learned video compression (LVC) have led to significant performance gains, with codecs such as DCVC-RT surpassing the H.266/VVC low-delay mode in compression efficiency. However, existing LVCs still exhibit key limitations: they often require separate models for intra and inter coding modes, and their performance degrades when temporal references are unreliable. To address this, we introduce Uni-LVC, a unified LVC method that supports both intra and inter coding with low-delay and random-access in a single model. Building on a strong intra-codec, Uni-LVC formulates inter-coding as intra-coding conditioned on temporal information extracted from reference frames. We design an efficient cross-attention adaptation module that integrates temporal cues, enabling seamless support for both unidirectional (low-delay) and bidirectional (random-access) prediction modes. A reliability-aware classifier is proposed to selectively scale the temporal cues, making Uni-LVC behave closer to intra coding when references are unreliable. We further propose a multistage training strategy to facilitate adaptive learning across various coding modes. Extensive experiments demonstrate that Uni-LVC achieves superior rate–distortion performance in intra and inter configurations while maintaining comparable computational efficiency.
\end{abstract}

\begin{IEEEkeywords}
Learned Image Compression, Learned Video Compression, Unified Method
\end{IEEEkeywords}

\section{Introduction}
\label{sec:intro}

Learned video compression (LVC) has recently achieved notable progress, surpassing traditional rule-based codecs such as H.266/VVC low-delay mode~\cite{bross2021overview} in rate-distortion performance~\cite{jia2025towards,jiang2025ecvc,zhang2025flavc,tang2025neural,bian2025augmented,tong2025context,phung2025mh,chen2025hytip,liao2025ehvc,jiang2025biecvc,sheng2025bi,liu2025neural}.

\begin{figure*}[t]
    \centering
    \subfloat[PSNR vs. frame index\label{fig:psnr}]{%
        \includegraphics[width=0.47\linewidth]{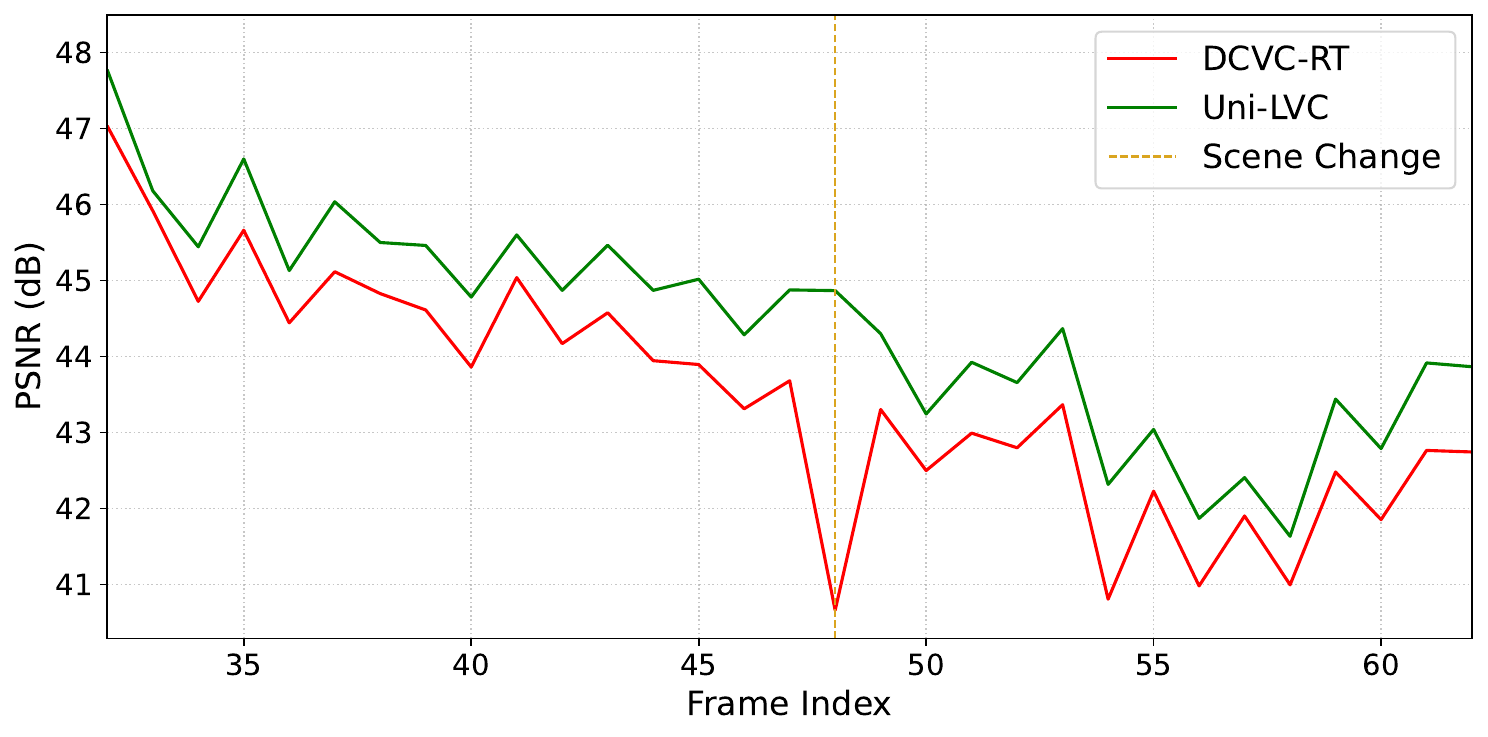}%
    }
    \hfill
    \subfloat[BPP vs. frame index\label{fig:bpp}]{%
        \includegraphics[width=0.47\linewidth]{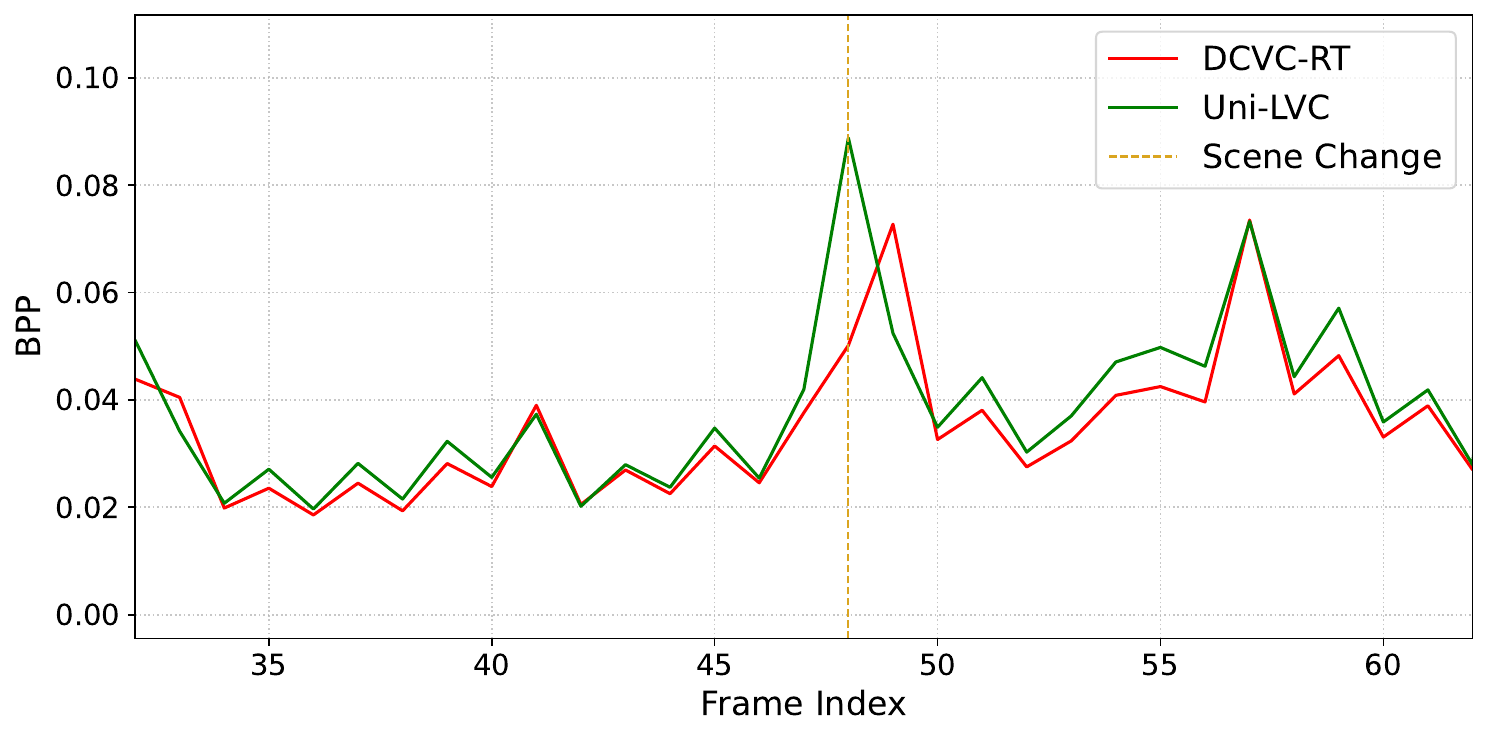}%
    }
    \caption{\textbf{PSNR and BPP vs frames.} 
The results are obtained on \texttt{videoSRC21} sequence from MCL-JCV, which contains a scene change at frame 48. 
At this frame, DCVC-RT continues to rely on previous time references, causing a sharp drop in PSNR. 
In contrast, Uni-LVC automatically suppresses unreliable temporal features and switches to intra‐dominant coding ($\alpha_t \approx 0.1$), keeping stable PSNR quality.
}
    \label{fig:scenechange}
\end{figure*}

Despite these advances, most existing approaches target narrow operating regimes: models are designed either for \emph{intra}~\cite{li2025learned,lu2025learned,feng2025linear} or for \emph{inter}~\cite{jia2025towards,jiang2025ecvc,liao2025ehvc}, and inter codecs are typically specialized for either low-delay (LD; unidirectional prediction)~\cite{jia2025towards,jiang2025ecvc,liao2025ehvc} or random-access (RA; bidirectional prediction)~\cite{jiang2025biecvc,sheng2025bi,liu2025neural} settings. This specialization complicates deployment and precludes seamless switching across modes in practical communication scenarios. Moreover, because inter-models rely heavily on temporal information, their performance often collapses at scene changes or when temporal references are corrupted, mismatched, or otherwise unreliable, as shown in~\Cref{fig:scenechange}.

In contrast, traditional hybrid video codecs such as H.265/HEVC~\cite{sullivan2012overview} and H.266/VVC~\cite{bross2021overview} inherently avoid these issues through their unified coding architecture. Hybrid codec standards are designed with a single coding pipeline that supports all coding modes, namely intra, inter (LD \& RA), and even mixed configurations, within the same framework. This is achieved by explicitly defining syntax and tools (e.g., CTUs~\cite{huang2021block}, reference frame lists, and GOP structures~\cite{chien2021motion,yang2021subblock}) that enable seamless switching between intra and inter predictions without requiring separate models or retraining. Moreover, the deterministic motion estimation and compensation processes in these codecs are hand-crafted and rule-based, allowing them to gracefully handle unreliable temporal references or scene changes by automatically inserting intra-refresh frames or reinitializing reference buffers based on rate-distortion optimization (RDO)~\cite{chien2021motion,yang2021subblock}. As a result, conventional codecs exhibit strong stability and robustness across diverse video content and transmission conditions.

Recent efforts partially bridge these gaps. Liu \textit{et al.}~\cite{liu20242} propose a diffusion-based perceptual coding framework unifying intra and inter within one model, but diffusion inference is costly, and their approach lacks mechanisms to mitigate unreliable temporal cues. Chen \textit{et al.}~\cite{chen2025hytip} improve reference quality via a hybrid referencing buffer scheme, yet the method remains low-delay only and still depends on accurate motion estimation. Jiang \textit{et al.}~\cite{jiang2025ecvc,jiang2025biecvc} exploit both local and non-local correlations from multiple references, but their models remain specialized for particular configurations rather than unified. Liao \textit{et al.}~\cite{liao2025ehvc} improve reference quality via a hierarchical multi-reference scheme, yet the method remains low-delay only and still depends on accurate motion estimation.

In this paper, we present {Uni-LVC}, a unified LVC method that supports both intra and inter (LD \& RA) coding modes in a single model. Uni-LVC treats inter-coding as intra-coding conditioned on reliable temporal features from reference frames. We first build a strong intra-codec backbone and introduce a lightweight cross-attention adaptation mechanism to efficiently integrate temporal cues. To ensure robustness, a reliability-aware temporal classifier dynamically regulates the use of temporal features to prevent performance degradation under poor references and challenging reference conditions. Finally, a progressive multistage training strategy enables effective adaptation across diverse coding modes. Across extensive experiments, Uni-LVC delivers superior rate-distortion performance in both intra and inter configurations while keeping runtime and complexity comparable to prior learned codecs.

Our main contributions are summarized as follows:
\begin{itemize}
    \item We propose {Uni-LVC}, a unified method that supports both intra and inter (LD \& RA) using a single model.
    \item We build a powerful intra-coding codec that surpasses prior learned intra codecs, serving as a solid foundation.
    \item We design a reliable temporal path with hybrid cross-attention and a reliability-aware classifier that adaptively suppresses unreliable temporal cues.
    \item We propose a multistage training scheme that enables effective learning across diverse coding modes (intra, LD, and RA) in a single model.
\end{itemize}

\section{Related work}
\label{sec:related}

\subsection{Learned image compression}
\label{sec:lic}

Learned image compression (intra coding) methods represent images as compact latents that are quantized and entropy coded. Early methods~\cite{minnen2018joint,balle2020nonlinear} parameterize the nonlinear transform function by CNNs and subsequent work explores attention mechanisms~\cite{cheng2020learned}, frequency-domain representations~\cite{li2024frequencyaware}, invertible or flow-based architectures~\cite{xie2021enhanced,cai2024i2c} and transformer architectures~\cite{koyuncu2022contextformer,liu2023learned,feng2025linear} and state-space models such as Mambas~\cite{qin2024mambavc,wu2025cmamba,zeng2025mambaic,qin2025cassic}, leading to notable improvements in compression efficiency. Additional advances include learnable quantization~\cite{li2025learnable}, codebook/vector quantization~\cite{zhu2022unified,xu2025multirate,qi2025generative}, dictionaries~\cite{lu2025learned}, optimization strategies~\cite{zhang2025balanced,li2025on,zhang2026leveraging} and hierarchical representations/ modellings~\cite{hu2020coarse,duan2023qarv,li2025learned,zhang2026qarv++}. 
Together, these developments provide a strong foundation for all intra coding.

\subsection{Learned video compression}
Extending intra coding to the temporal domain, learned video compression exploits temporal redundancy via motion estimation/compensation and temporal context learning. Early approaches~\cite{lu2019dvc,agustsson2020scale,lin2020m,hu2022coarse,rippel2021elf} follow the classical residual-coding paradigm, transmitting the prediction error between the current frame and the motion-compensated reference. DCVC~\cite{li2021deep} shifts to contextual conditional coding, conditioning temporal features on the current frame to learn dependencies adaptively. Following DCVC, numerous extensions have further advanced temporal modeling and efficiency. DCVC-TCM~\cite{sheng2022temporal} incorporates multi-scale temporal context modeling, while DCVC-HEM~\cite{li2022hybrid} introduces multi-granularity quantization and dual spatial entropy models for variable-rate control. DCVC-DC~\cite{li2023neural} enhances context diversity through offset-based modeling and introduces quadtree partitioning for the entropy model, and DCVC-FM~\cite{li2024neural} extends the framework to support a wide range of bitrates and introduces periodic refresh mechanisms to mitigate temporal error propagation. DCVC-RT~\cite{jia2025towards} optimizes the architecture for real-time encoding/decoding with competitive performance. Transformer-based codecs such as ECVC/BiECVC~\cite{jiang2025ecvc,jiang2025biecvc} model local/global dependencies from multiple references. EHVC~\cite{liao2025ehvc} aggregates from preceding and key frames, and DCVC-B~\cite{sheng2025bi}, BRHVC~\cite{liu2025neural} enable bidirectional prediction for RA. Practical deployment~\cite{van2024mobilenvc,gao2025pnvc,chen2025learning} and hybrid buffering strategies~\cite{chen2025hytip} have also been studied.

However, as discussed above, most LVCs remain specialized for a single mode (intra vs.\ inter) or a single inter configuration (LD vs.\ RA), complicating switching and deployment. Moreover, inter codecs typically consume temporal features regardless of their quality, risking severe degradation under corrupted or mismatched references. These challenges motivate a unified framework that supports all coding modes and adaptively regulates temporal information.

\section{Proposed method}
\subsection{Overview}
\label{sec:overview}

\begin{figure}[t]
    \centering
    \includegraphics[width=\linewidth]{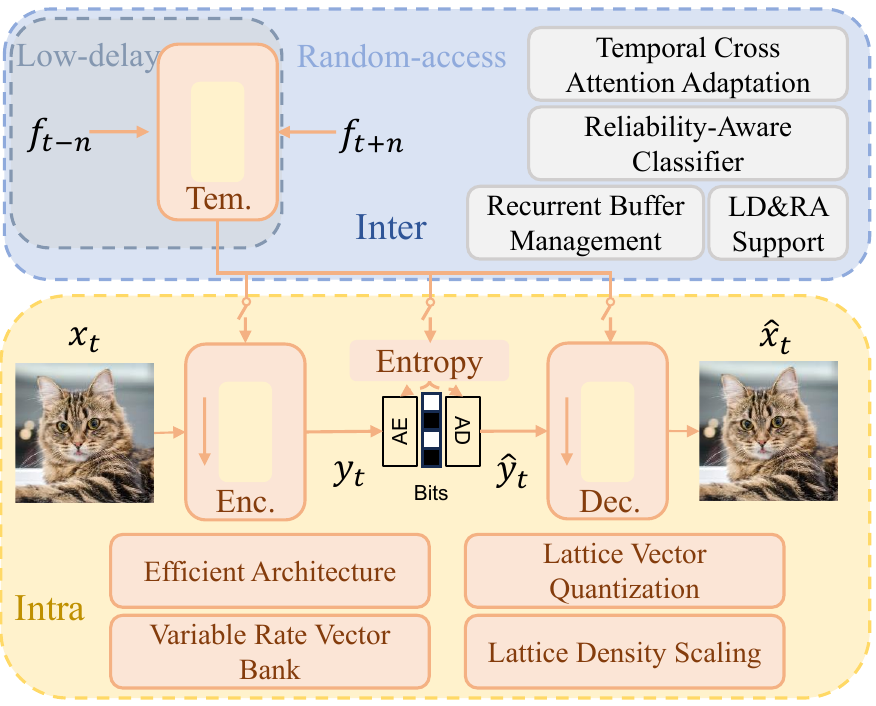}
    \caption{
        Overview of the proposed {Uni-LVC}. 
        ``Enc.'' and ``Dec.'' denote the encoder and decoder, respectively, while ``Tem.'' represents the temporal modeling module. 
        Inter coding (LD and RA) is formulated as intra coding conditioned on auxiliary temporal features extracted from temporal references stored in the buffer.
    }
    \vspace{-0.6cm}
    \label{fig:overview}
\end{figure}

An overview of the proposed {Uni-LVC} method is illustrated in~\Cref{fig:overview}. 
Uni-LVC is designed to unify intra and inter video coding in a single model by treating inter coding as intra coding conditioned on temporal information. 
The intra codec compresses each frame independently through an encoder–decoder with entropy modeling, following standard practice~\cite{balle2020nonlinear}. 
For inter coding, temporal references from previously decoded unidirectional (LD) or bidirectional (RA) frames are processed by a temporal modeling module to adaptively extract reliable auxiliary temporal features. 
These features are then integrated into the intra codec through a cross-attention adaptation mechanism, enabling the model to reduce temporal redundancy without altering the underlying intra network. Through this unified design, Uni-LVC flexibly supports all intra (AI), LD, and RA configurations in a single model, enabling efficient and robust learned video compression across diverse scenarios.

\subsection{Intra coding}
\label{sec:intra}
A strong intra backbone is essential for unified coding across modes. We thus adopt DCVC-RT~\cite{jia2025towards} as the baseline and strengthen it with a simplified hierarchical progressive context model (HPCM)~\cite{li2025learned} and a learned lattice vector quantizer~\cite{xu2025multirate}, improving entropy modeling and quantization efficiency while retaining practical complexity.

\begin{figure}[htbp]
    \centering
    \includegraphics[width=0.9\linewidth]{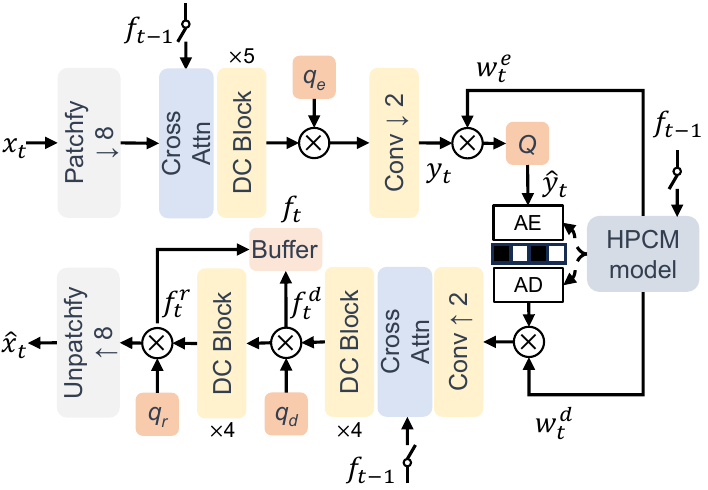}
    \caption{{Architecture of the proposed intra codec.} ``Cross Attn'' and ``DC Block'' denote the cross-attention module and enhanced depthwise convolution block. ``Q'' indicates quantization. A hybrid buffer $f_t$ is formed from $f_t^d$ and $f_t^r$ and stores temporal references. For inter coding, the temporal feature $f_{t-1}$ (or its bidirectional counterpart) is fed through Cross Attn; for pure intra coding, $f_{t-1}$ is a quality-specific learnable vector $f_p$ to preserve compatibility and learn global priors. HPCM denotes the hierarchical progressive context model.}
    \label{fig:intra}
\end{figure}

\subsubsection{Neural architecture}
The architecture of the proposed main intra codec is illustrated in~\Cref{fig:intra}. 
Following DCVC-RT~\cite{jia2025towards}, the input frame is first converted into a low-resolution representation via a pixel-unshuffle~\cite{shi2016real}, achieving an $8\times$ spatial downsampling. 
A lightweight cross-attention is then applied to optionally fuse temporal reference features from the buffer, enabling seamless reuse of the same architecture for both intra and inter coding. Note for intra coding, no temporal information is given. $f_{t-1}$ is a quality-specific learnable vector $f_p$.
The core of the encoder and decoder is a stack of enhanced depthwise convolution (DC) blocks, which provide a favorable balance between efficiency and modeling capacity. 
At the end of the encoder, a strided convolution further downsamples the latent by a factor of~2, while the decoder performs the symmetric upsampling via a convolution+pixel-shuffle~\cite{shi2016real}. 
Learnable rate-control vectors $q$ are injected into both encoder and decoder, enabling variable-rate operation (details in~\Cref{sec:Variable}). 
For entropy modeling, we adopt the simplified hierarchical progressive context model (HPCM)~\cite{li2025learned}, and apply lattice vector quantization~\cite{xu2025multirate} to the latent space due to its superior space-filling efficiency and coding performance.

\begin{figure}[htbp]
    \centering
    \includegraphics[width=\linewidth]{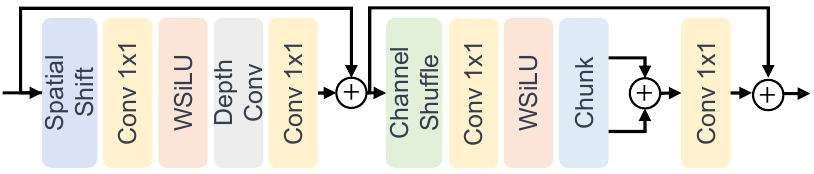}
    \caption{
Enhanced depthwise convolution (DC) block.
}
    \label{fig:edc}
\end{figure}
\textbf{Enhanced depthwise convolution block.}
We build on the DC block from DCVC-RT~\cite{jia2025towards}, where depthwise convolution handles spatial feature extraction and an MLP performs channel mixing. 
The weighted SiLU activation (WSiLU)~\cite{elfwing2018sigmoid} is adopted, defined as $\mathrm{WSiLU}(x) = x \cdot \sigma(4x)$. ~\Cref{fig:edc} illustrates our enhanced block.

To strengthen spatial interaction beyond the original design, we prepend a parameter-free spatial-shift operation~\cite{yu2022s2} to the block. 
The first half of channels is evenly split into four groups and shifted by one pixel along $\pm x$ and $\pm y$ (right, left, down, up), while the remaining channels are left unchanged, allowing each patch to absorb visual context from its neighbors without introducing extra parameters or computation. 
When stacked across multiple blocks, the shifted features gradually diffuse over the entire spatial domain, enhancing spatial coupling while preserving the lightweight nature of the architecture.

To improve channel interaction, we further insert a channel shuffle operation~\cite{zhang2018shufflenet} before the MLP. 
Channel shuffle permutes channel groups so that features mixed by the MLP are redistributed in subsequent blocks, enabling cross-channel communication at zero additional cost. Together, the spatial shift, depthwise convolution, channel shuffle, and MLP form an efficient spatio–channel mixing unit.

\begin{figure}[htbp]
    \centering
    \includegraphics[width=1\linewidth]{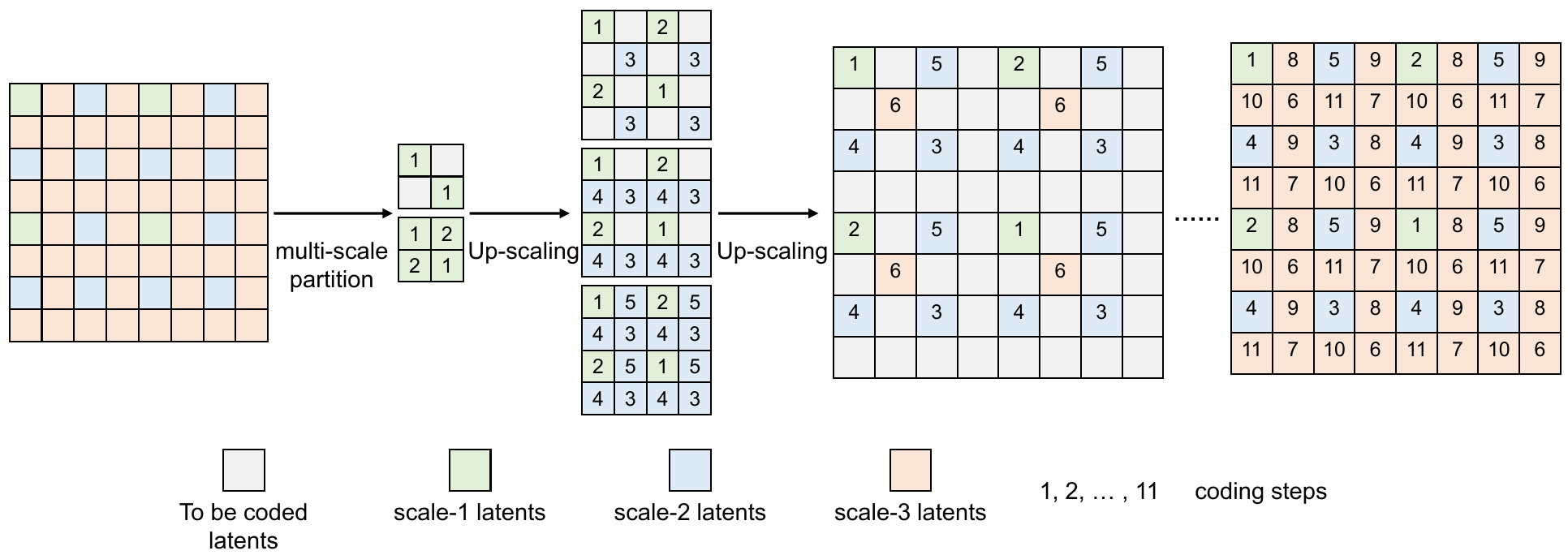}
    \caption{
        Hierarchical progressive context model.
    }
    \label{fig:hpcm}
\end{figure}

\textbf{Hierarchical progressive context model (HPCM).}
We adopt the HPCM~\cite{li2025learned} as our entropy model.
HPCM progressively refines probability estimation through multi-stage context aggregation, enabling accurate entropy modeling with low computational cost.
Our implementation follows the original structure and prediction pipeline, with one simplification: we remove channel grouping to better integrate with lattice vector quantization, rendering the model a purely spatial autoregressive coder. Given the quantized latent $\hat{y}$ and the hyperprior $h_s(\hat{z})$\footnote{The hyperprior also predicts $w^e$/$w^d$, which are the encoder/decoder element-wise scaling parameters before and after quantization.}, HPCM produces the mean and scale parameters for the generalized Gaussian prior at each coding step:
\[
(\mu_i,\sigma_i) = \mathrm{HPCM}\!\left(\hat{y}_{<i},\, h_s(\hat{z})\right),
\]
where $\hat{y}_{<i}$ denotes all latents that have already been coded before step $i$.
As illustrated in~\Cref{fig:hpcm}, we partition $\hat{y}$ on the spatial grid into three nested subsets at different scales,
$\hat{y}^{S_1}$, $\hat{y}^{S_2}$ and $\hat{y}^{S_3}$.
$\hat{y}^{S_1}$ is obtained by regular subsampling with stride $4$ (coarsest scale),
$\hat{y}^{S_2}$ refines it with additional positions at stride $2$,
and $\hat{y}^{S_3}$ corresponds to the full-resolution latent grid.
The three scales are coded in a coarse-to-fine manner:
we first code all elements of $\hat{y}^{S_1}$, then $\hat{y}^{S_2}$, and finally $\hat{y}^{S_3}$.
After finishing a scale, the decoded symbols are written back and up-scaled to the next scale so that every later step can condition on all previously decoded coarser-resolution latents.
This results in $11$ coding steps in total (2 on $S_1$, 3 on $S_2$, and 6 on $S_3$) as shown in~\Cref{fig:hpcm}.

\textbf{Lattice Vector Quantization.}
We employ a learned lattice vector quantizer (LVQ)~\cite{xu2025multirate,zhang2024learning}, which jointly quantizes groups of latent elements rather than treating them independently, improving R-D efficiency by increasing packing density in the latent space and reducing redundancy.

Let the encoder produce a latent representation $y \in \mathbb{R}^n$. A \emph{lattice} $\Lambda \subset \mathbb{R}^n$ is defined as the set of all integer linear combinations of $n$ linearly independent basis vectors $\{b_1,\dots,b_n\}$:
\begin{equation}
\small
\Lambda = \{ z \mid z = Bu,\; u \in \mathbb{Z}^n \}
= \left\{ \sum_{i=1}^n u_i b_i \;\middle|\; u_i \in \mathbb{Z} \right\},
\end{equation}
where $B = [b_1,\dots,b_n] \in \mathbb{R}^{n \times n}$ is the \emph{lattice basis matrix}. In an LVQ, the codebook is given by the lattice points of~$\Lambda$.

To adapt the lattice geometry to the data distribution, the basis matrix $B$ is learned during training. We parameterize $B$ via its singular value decomposition (SVD):
\(
B = U \Sigma V^{\top},
\)
where $U$ and $V$ are learnable orthogonal matrices, and $\Sigma$ is a diagonal matrix with non-zero entries, ensuring invertibility. Orthogonality is enforced via PyTorch’s built-in orthogonal parameterization.

Given the quantized latents $\hat{y}$, we model the latents with a generalized Gaussian distribution
$\mathcal{N}_{\beta}(\mu, \sigma)$, where $\mu$ and $\sigma$ are predicted by the
entropy model and the shape parameter $\beta$ is a learnable model-wise parameter~\cite{zhang2025generalized, li2025learned}. Different $\beta$ are learned for different modes (AI, LD, and RA) due to their unique characteristics. The latent ${y}$ is first transformed into
the lattice coordinate system via
\(
y_{tr} = B^{-1}(y - \mu),
\)
and quantized using element-wise rounding
(during training, we employ straight-through estimator (STE) and additive uniform noise~\cite{liu2023learned}),
\(
\hat{y}_{tr} = \mathrm{round}(y_{tr}),
\)
which corresponds to Babai’s nearest-plane method~\cite{babai1986lovasz}, an efficient approximation to nearest-lattice decoding.

The quantized coefficients are entropy modeled by:
\begin{equation}
\small
p(\hat{y}_{tr}) = \prod_{i=1}^{n}
\Big[
\big(\mathcal{N}_{\beta}(0, \sigma_i) * \mathcal{U}(-\tfrac{1}{2}, \tfrac{1}{2})\big)
\big(\hat{y}_{tr}^{(i)}\big)
\Big],
\end{equation}
where the generalized Gaussian is convolved with a uniform distribution. At the decoder side, the inverse lattice transform recovers the latent
representation,
\(
\hat{y} = B \hat{y}_{tr} + \mu.
\)

\subsubsection{Variable rate coding}
\label{sec:Variable}

To support practical usage under diverse bitrate constraints, the proposed codec enables variable-rate through: \emph{learned rate-control vectors} and \emph{lattice density scaling}.

\textbf{Rate-control vector banks.}
Following~\cite{jia2025towards}, we adopt learnable vector banks to adapt to the bitrate. 
For each quality level, a learnable vector scales the feature.
Separate banks are maintained for different modules: the encoder ($q_e$), decoder ($q_d$), feature buffer ($q_f$), and reconstruction head ($q_r$). In addition, element-wise scaling parameters $w_t^e$ and $w_t^d$, predicted by the hyperprior model, are applied before and after quantization for fine-grained control.

\textbf{Lattice density scaling.}
We also control the bitrate by adapting the density of the learned lattice quantizer. 
The density is adjusted by a scalar factor $a$: increasing $a$ yields a denser lattice, reducing quantization error and improving reconstruction quality at the cost of a higher bitrate. 
Rather than modifying the lattice basis directly, we adopt an equivalent formulation: the latent $y$ is first scaled by $a$ before quantization, and the decoded coefficients are multiplied by $\tfrac{1}{a}$ to restore the original scale. 
Each quality level is associated with a learnable $a$ value, initialized following~\cite{tong2023qvrf} and optimized jointly with the codec.

\subsection{Inter coding}
Building on the solid foundation of the proposed intra codec, Uni-LVC performs inter coding by seamlessly injecting previous state temporal references into the same model by the cross-attention block in~\Cref{fig:intra}. 

\subsubsection{Buffer management}
To support temporal referencing in inter coding, we maintain a buffer that stores hybrid features derived from previously decoded frames. 

\begin{figure}[htbp]
    \centering
    \includegraphics[width=0.75\linewidth]{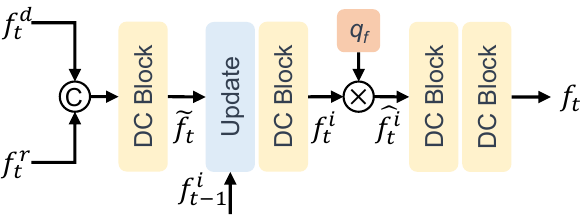}
    \caption{
        Buffer feature extraction.
    }
    \label{fig:buffter}
\end{figure}

As illustrated in ~\Cref{fig:intra} and ~\Cref{fig:buffter}, the buffer feature is generated from a hybrid input composed of the decoder feature $f_t^{d}$ and the reconstruction feature $f_t^{r}$. 
DC blocks and a recurrent update mechanism produce the stored feature $f_t^{i}$.

\textbf{Recurrent update mechanism.}
The $f_t^{d}$ and $f_t^{r}$ are first concatenated and merged by a DC block to generate a feature $\tilde{f}_t$ for the current timestep.
To retain useful long-range temporal cues, the buffer is implemented as a recurrent state (Update block in~\Cref{fig:buffter}).  
The $\tilde{f}_t$ is fused with the previous stored feature $f^{i}_{t-1}$ via a gated update (LSTM-style~\cite{beck2024xlstm}):
\[
\small
\begin{aligned}
g_t^{\mathrm{F}} &= \sigma\!\left(\mathrm{DC}_{\mathrm{F}}\!\big([f^{i}_{t-1},\, \tilde{f}_t]\big)\right), \
g_t^{\mathrm{I}} = \sigma\!\left(\mathrm{DC}_{\mathrm{I}}\!\big([f^{i}_{t-1},\, \tilde{f}_t]\big)\right), \\
&\hspace{4.5em} f^{i}_{t} = g_t^{\mathrm{F}} \odot f^{i}_{t-1} \;+\; g_t^{\mathrm{I}} \odot \tilde{f}_t .
\end{aligned}
\]
where $[\cdot,\cdot]$ denotes concatenation, $\sigma(\cdot)$ is the sigmoid, and $\odot$ is element-wise multiplication.
This update enables the buffer to selectively retain long-range history while incorporating new temporal evidence.

When temporal features are needed for inter coding, the updated stored feature $f^{i}_{t}$ is then modulated by a learnable rate-control vector, \(
\hat{f}^{i}_{t} = q_f \odot f^{i}_{t},
\)
to adjust the contribution of the buffer and ensure full compatibility with variable-rate coding.
Two DC blocks then transform $\hat{f}^{i}_{t}$ into the temporal feature $f_t$ for later usage. The stored state is $f^{i}_{t}$; the rate-adapted output for usage is $f_t$.

\textbf{Bidirectional feature.}
For RA, both past (backward) and future (forward) references relative to the current timestep $t$ are available. 
Accordingly, the buffer maintains two states: a backward state $f^{i,\mathrm{bwd}}_{t}$ and a forward state $f^{i,\mathrm{fwd}}_{t}$.
Each state is converted to a rate-adapted temporal feature using the same pathway:
\[\small
\begin{aligned}
f^{\mathrm{bwd}}_{t} = \mathrm{DC}\left(\mathrm{DC}\!\left(q_f \odot f^{i,\mathrm{bwd}}_{t}\right)\right),
f^{\mathrm{fwd}}_{t} = \mathrm{DC}\left(\mathrm{DC}\!\left(q_f \odot f^{i,\mathrm{fwd}}_{t}\right)\right).
\end{aligned}\] To provide a unified representation to the inter-prediction module, the forward and backward features are fused by concatenation followed by a DC block:
\[\small
\begin{aligned}
f^{\mathrm{bi}}_{t} = \mathrm{DC}_{\mathrm{merge}}\!\big([\,f^{\mathrm{bwd}}_{t},\, f^{\mathrm{fwd}}_{t}\,]\big),
\end{aligned}
\]
yielding a single feature $f^{\mathrm{bi}}_{t}$ that matches the dimensionality of the unidirectional buffer feature $f_t$. 
This allows the temporal adaptation to operate with the same model architecture in both modes; the differences are whether one or two features are retrieved, and for the RA buffer, the update block is skipped due to the complexity of RA hierarchical layered structure~\footnote{Only the previous instant state is utilized without long-term memory}.

\subsubsection{Classifier guidance}
\label{sec:CG}
Although the buffer provides temporal cues, their usefulness is not guaranteed: scene cuts, motion discontinuities, or corrupted references may introduce misleading features that can degrade inter prediction.  
To address this, we introduce a lightweight reliability-aware classifier that adaptively regulates the contribution of the temporal feature.

Given the current input frame $x_t$ and the ready to use temporal feature ($f_{t-1}$ for LD or $f^{\mathrm{bi}}_{t-1}$ for RA), we compute
\begin{equation}
\small
\alpha_t=\sigma\!\Big(\mathrm{GAP}\big(\mathrm{Cls}([x_t,\,f_{t-1}^{\ast}])\big)\Big),\
f_{t-1}^{\ast}\in\{f_{t-1},\,f^{\mathrm{bi}}_{t-1}\},
\end{equation}
where $\mathrm{Cls}(\cdot)$ is a two-layer DC block, $\mathrm{GAP}(\cdot)$ denotes global average pooling, and $\sigma(\cdot)$ is the sigmoid.  
The $\alpha_t \in [0,1]$ is a single scalar that reflects the global reliability of the temporal reference: values near $1$ indicate high-confidence temporal cues, while values near $0$ suppress them. The temporal feature is then globally gated by: \(\alpha_t \cdot f_{t-1}^{\ast},
\)
and forwarded to the cross-attention module. Final utilized feature is $\alpha_t \cdot f_{t-1}^{\ast} + f_p$ for inter to support intra style when $f_{t-1}^{\ast}$ is gated with low reliability $\alpha_t$. $\alpha_t$ is packed into the bitstream as side information with 16-bit precision in the header, which is negligible per frame.

\subsubsection{Cross-attention adaptation}
\label{sec:CA}
To exploit temporal redundancy while preserving a unified intra-inter architecture, Uni-LVC injects reference into the current coding step through an \emph{efficient hybrid cross-attention module} (Cross Attn), treating inter-coding as intra-coding \emph{conditioned} on temporal cues, rather than relying on a separate motion compensation. Our Cross Attn consists of two branches: {Deformable Neighborhood Cross-Attention (DN-CA)} for \emph{local} correspondence, and {Polarity-Aware Linear Cross-Attention (PAL-CA)} for \emph{global} temporal interactions with linear complexity. The outputs of the two branches are summed, enabling each query to attend to a deformable local neighborhood and a full-frame temporal context. Cross Attn is inserted into the encoder, decoder, and entropy model so that temporal information influences feature encoding/decoding and probability estimation in a unified manner, and it utilizes decoded references to ensure decoder-consistent conditioning.

\begin{figure}[htbp]
    \centering
    \includegraphics[width=0.9\linewidth]{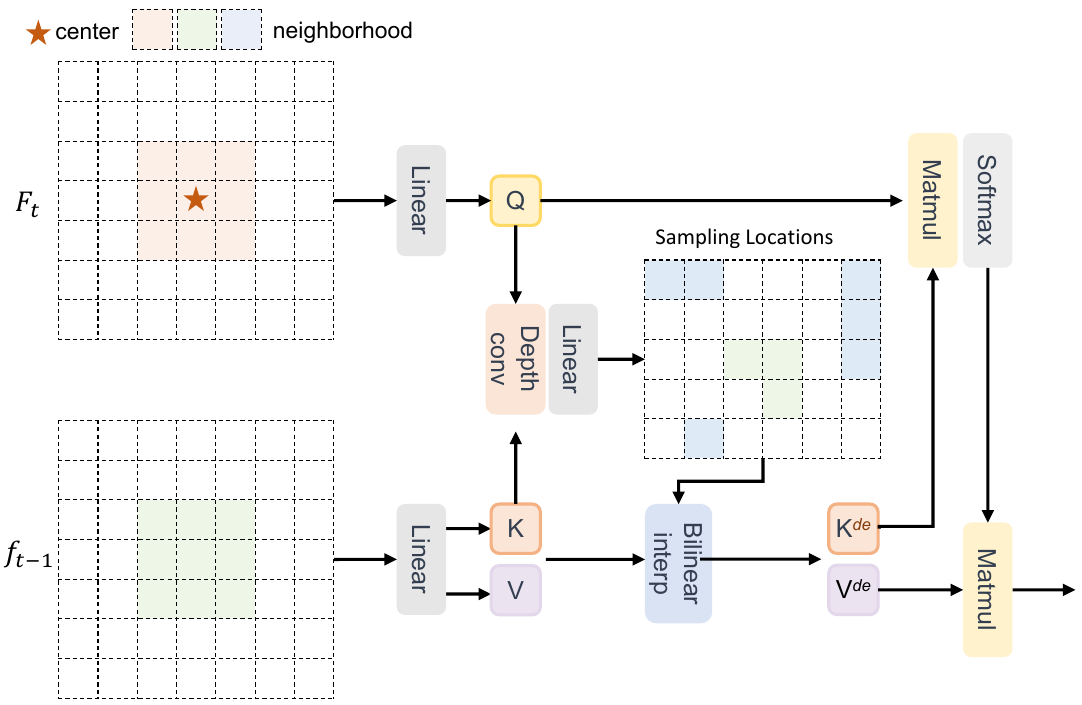}
    \caption{
        Deformable neighborhood cross-attention.
    }
    \label{fig:dnca}
\end{figure}

\textbf{Deformable neighborhood cross-attention (DN-CA).}
Given the current feature $F_t$ and the temporal reference $f_{t-1}$, query, key, and value tensors are generated:
\[
\small
\begin{aligned}
Q_t = W_Q F_t,\ 
K_{t-1} = W_K f_{t-1},\ 
V_{t-1} = W_V f_{t-1}.
\end{aligned}
\]
For each spatial location $(i,j)$, the colocated query and key vectors are concatenated and fed into a lightweight offset-prediction network $f_\theta$ (a depthwise $3\times3$ convolution followed by a linear layer), which outputs $k^2$ 2D offsets, where $k=5$ is the neighborhood size:
\[
\small
\begin{aligned}
\Delta p_{i,j} = f_\theta([q_{i,j},k_{i,j}]) \in \mathbb{R}^{k^2 \times 2},
\end{aligned}
\]
defining a deformable sampling neighborhood $\Omega_k(i,j)$~\cite{hassani2023neighborhood}. Keys and values at these positions are obtained by bilinear interpolation~\cite{xiong2024efficient}:
\[
\small
\begin{aligned}
k_{i,j}^{(m)} = \mathrm{BiSample}\!\left(K_{t-1},\, p_{i,j}^{(m)}\right),\
v_{i,j}^{(m)} = \mathrm{BiSample}\!\left(V_{t-1},\, p_{i,j}^{(m)}\right),
\end{aligned}
\]
where $p_{i,j}^{(m)}$ is the $m$-th deformed location in $\Omega_k(i,j)$. A standard dot-product attention is then computed within this neighborhood to produce the local output:
\begin{equation}
\small
\label{eq:local_attn_final}
Z_{i,j}^{(\mathrm{local})} =
\sum_{m \in \Omega_k(i,j)}
\mathrm{Softmax}_{m}
\!\left(
    \frac{q_{i,j} \cdot k_{i,j}^{(m)}}{\sqrt{d_k}}
\right)
v_{i,j}^{(m)}.
\end{equation}
All operations are applied in a per-head manner. The detailed architecture is illustrated in~\Cref{fig:dnca}.

\begin{figure}[htbp]
    \centering
    \includegraphics[width=0.7\linewidth]{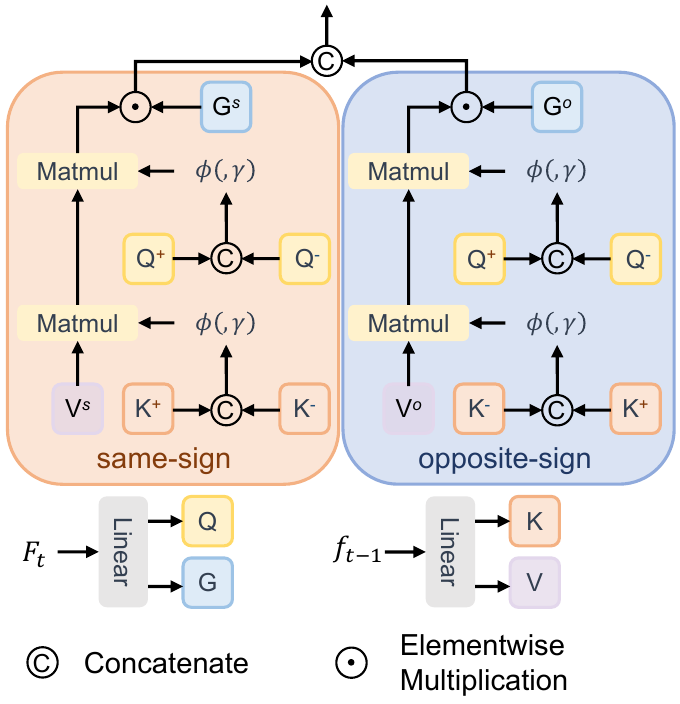}
    \caption{
        Polarity-aware linear cross-attention.
    }
    \label{fig:palca}
\end{figure}
\textbf{Polarity-aware linear cross-attention (PAL-CA).}
While DN-CA captures local motion, it cannot model long-range temporal dependencies such as large camera motion. To complement it, we introduce a global branch based on linear attention~\cite{meng2025polaformer}.
Two linear layers applied to $F_t$ and $f_{t-1}$ produce the global query $Q_g$, key $K_g$, value $V_g$, and an additional gating tensor $G$ for modulating the outputs (\Cref{fig:palca}):
\[
\small
\begin{aligned}
Q_g = W_{Q_g} F_t,\quad 
K_g = W_{K_g} f_{t-1},\quad 
V_g = W_{V_g} f_{t-1}.
\end{aligned}
\]
Following PolaFormer~\cite{meng2025polaformer}, $Q_g$ and $K_g$ are decomposed into positive and negative parts to separate constructive and destructive correlations:
\[
\small
\begin{aligned}
Q_g^{+} &= \mathrm{ReLU}(Q_g), \qquad
Q_g^{-} = \mathrm{ReLU}(-Q_g), \\
K_g^{+} &= \mathrm{ReLU}(K_g), \qquad
K_g^{-} = \mathrm{ReLU}(-K_g),
\end{aligned}
\]
and $V_g$ is split along the channel dimension into $V^s$ (same-sign path) and $V^o$ (opposite-sign path). All polarity components are passed through a learnable power function
$\phi(x) = x^{\gamma}$,
where $\gamma$ is a channel-wise learnable parameter. Let
\[
\small
\begin{aligned}
\phi_Q = \phi([Q_g^+,Q_g^-]),\
\phi_{K_s} = \phi([K_g^+,K_g^-]),\
\phi_{K_o} = \phi([K_g^-,K_g^+]).
\end{aligned}
\]
Using the kernel rearrangement trick for linear complexity, the same-sign and opposite-sign outputs are computed:
\[
\small
\begin{aligned}
O^s = 
\frac{\phi_Q \big(\phi_{K_s}^{\top} V^s\big)}
     {\phi_Q \big(\phi_{K_s}^{\top} \mathbf{1}\big)},\
O^o = 
\frac{\phi_Q \big(\phi_{K_o}^{\top} V^o\big)}
     {\phi_Q \big(\phi_{K_o}^{\top} \mathbf{1}\big)}.
\end{aligned}
\]
Two learnable coefficient matrices $G^s$ and $G^o$, obtained from $G$, gate the corresponding outputs, and the global response is formed by channel-wise concatenation:
\begin{equation}
\small
Z^{(\mathrm{global})} = 
[G^s \odot O^s,\; G^o \odot O^o].
\end{equation}

\textbf{Cross-attention block.}
The final Cross Attn output is obtained by summing the local and global branches:
\begin{equation}
\small
Z = Z^{(\mathrm{local})} + Z^{(\mathrm{global})},
\end{equation}
which is then added back to the input feature via a residual connection in the backbone. DN-CA provides precise deformable local matching, while PAL-CA supplies full-frame temporal reasoning with linear complexity, jointly enabling robust temporal feature conditioning without modifying the intra architecture and effectively turning inter coding into conditional intra coding. Triton kernels are implemented for the attention computation.

\subsection{Training strategy}
Training the Uni-LVC model involves a multi-stage curriculum designed to optimize the unified architecture for its diverse coding modes. We first train the intra codec in isolation and then introduce temporal components in a controlled manner, progressing from LD to RA. Throughout, we use \emph{knowledge replay} to prevent catastrophic forgetting of previously learned modes~\cite{duan2024towards}.

\subsubsection{Intra-codec training}
\label{sec:train_intra}
We train the intra codec in two stages: (1) learn a single high–quality \emph{anchor} model, and (2) expand it to a variable-rate model. This stabilizes joint learning of the transform, hyperprior, LVQ, and HPCM, and avoids brittle behavior across rates~\cite{presta2025stanh,li2024frequencyaware}.

\textbf{Stage 1: Anchor model (single rate).}
We begin by training a high-quality anchor at the largest $\lambda$ (highest target quality, highest rate), and later use it to fine-tune a variable-rate codec. We adopt a simple two-step schedule. Note that, during this stage, the lattice basis $B$ and the rate-control vectors corresponding to the anchor quality are also optimized, providing a strong single-rate baseline.

\emph{Step 1: Transform \& hyperprior pretraining.}
We train the encoder and decoder together with the hyperprior encoder/decoder while \emph{disabling} HPCM. The goal is to learn a strong latent representation before introducing the more expressive entropy model. We optimize the standard rate–distortion objective
\begin{equation}
\small
\mathcal{L} \;=\; \mathcal{R}(\hat{y}) \;+\; \mathcal{R}(\hat{z}) \;+\; \lambda\,\mathcal{D}(x,\hat{x}),
\label{eq:rd_loss}
\end{equation}
where $\mathcal{D}$ is the distortion (e.g., MSE), and $\mathcal{R}(\hat{y})$/$\mathcal{R}(\hat{z})$ are the expected rates of the latents and hyper-latents, respectively, estimated from the hyperprior.

\emph{Step 2: Joint training with HPCM.}
We initialize from Step~1, enable HPCM (randomly initialized), and jointly train the entire model (transform, hyperprior, HPCM, and LVQ) under the same loss in \Cref{eq:rd_loss}, now computing $\mathcal{R}(\hat{y})$ using HPCM’s refined probability estimates.

\textbf{Stage 2: Variable-rate model.}
Starting from the converged anchor, we activate all variable-rate mechanisms from \Cref{sec:Variable}: the rate-control vector banks $(q_e,q_d,q_f,q_r)$ and the lattice density scalars $\{a\}$. Entries associated with the anchor quality inherit their learned values; the remaining entries are initialized following~\cite{tong2023qvrf}. At each iteration, we \emph{sample} a target quality (and its $\lambda$), select the corresponding vectors and $a$, and perform a forward/backward pass with the R–D loss in \Cref{eq:rd_loss}. To balance performance across operating points, we adopt a multi-objective training strategy~\cite{li2025learnable,zhang2025balanced,liu2023famo}, which mitigates gradient conflicts between rates.

\subsubsection{Inter codec training}
\label{sec:train_inter}
Building on the trained intra codec, we extend to inter coding in two {stages}: (1) LD with unidirectional reference, then (2) RA with bidirectional references. In both stages, we apply \emph{knowledge replay} by sampling the operating mode at each iteration so earlier modes remain well-optimized.

\textbf{Stage 1: Low-delay adaptation.}
We train on sequential clips with a single decoded reference in the buffer.

\emph{Mode sampling with knowledge replay.}
At each iteration, we sample a Bernoulli switch $s\!\in\!\{0,1\}$:
\[
\small
s=0 \Rightarrow \text{AI (all-intra)},\qquad
s=1 \Rightarrow \text{LD (low-delay)}.
\]
When $s{=}0$, temporal is off; when $s{=}1$, the buffer and temporal path are on. We begin with a higher AI probability and anneal toward a balanced ratio during training. Training windows start with an I-frame to warm up the buffer; gradients flow through the temporal path when $s{=}1$. Hierarchical quality and QP offset~\cite{jia2025towards,li2024neural} are enabled.

\emph{Losses and classifier calibration.}
For a training window $\mathcal{T}$, we minimize the rate–distortion objective from \Cref{eq:rd_loss} (with the same quantization relaxations), summed over $t\!\in\!\mathcal{T}$. To discourage gratuitous temporal usage and stabilize frame-to-frame behavior of the reliability gate $\alpha_t$ (defined in \Cref{sec:CG}), we add a sparsity-and-smoothness term~\cite{rudin1992nonlinear}:
\begin{equation}
\small
\mathcal{L}_{\mathrm{reg}}
=\sum_{t\in\mathcal{T}}\alpha_t
+\sum_{t\in\mathcal{T}}\lvert \alpha_t-\alpha_{t-1}\rvert.
\end{equation}
We supervise the reliability-aware classifier via positives and synthetically constructed negatives. Positives assume reliable temporal cues in clean clips ($r_t{=}1$). Negatives ($r_t{=}0$) are created by contaminating the temporal pathway with: (i) reference corruption (noise, blur, heavy compression)~\cite{salman2020adversarially,chen2023toward}, (ii) buffer replacement with low-information signals (e.g., solid colors), or (iii) mismatched buffers sampled from other sequences to emulate scene cuts. The gate is trained with label-smoothed BCE~\cite{szegedy2016rethinking},
\begin{equation}
\mathcal{L}_{\mathrm{cls}}
= \sum_{t\in\mathcal{T}}
\Big[
-\tilde{r}_t \log \alpha_t
-(1-\tilde{r}_t)\log(1-\alpha_t)
\Big],
\end{equation}
where $\tilde{r}_t = (1-\varepsilon)\,r_t + \varepsilon(1-r_t)$. The total LD loss is
\begin{equation}
\small
\mathcal{L}_{\mathrm{LD}}
=\sum_{t\in\mathcal{T}}\!\big(\mathcal{R}(\hat{y}_t)+\mathcal{R}(\hat{z}_t)+\lambda\,\mathcal{D}(x_t,\hat{x}_t)\big)
+\mathcal{L}_{\mathrm{cls}}
+\mathcal{L}_{\mathrm{reg}}.
\end{equation}

\textbf{Stage 2: Random-access adaptation.}
After LD converges, we enable bidirectional buffering and extend to RA. RA hierarchical quality and QP offset follow~\cite {sheng2025bi}.

\emph{Ternary mode sampling with knowledge replay.}
We sample a switch $u\!\in\!\{0,1,2\}$:
\[
\small
u=0 \Rightarrow \text{AI},\qquad
u=1 \Rightarrow \text{LD},\qquad
u=2 \Rightarrow \text{RA}.
\]
We begin with a small $p(u{=}2)$ and ramp it up, while keeping nonzero $p(u{=}0)$ and $p(u{=}1)$ to preserve AI/LD.

\emph{Losses and RA-specific negatives.}
We reuse the rate–distortion objective, the gate regularizer, and classifier supervision with additional RA-specific negatives: mixing forward/backward buffers across different sequences, or replacing one direction with corrupted features. The RA objective mirrors the LD:
\begin{equation}
\small
\mathcal{L}_{\mathrm{RA}}
=\sum_{t\in\mathcal{T}}\big(\mathcal{R}(\hat{y}_t)+\mathcal{R}(\hat{z}_t)+\lambda\,\mathcal{D}(x_t,\hat{x}_t)\big)
+\mathcal{L}_{\mathrm{cls}}+\mathcal{L}_{\mathrm{reg}}.
\end{equation}

Our schedule is: LD stage $(p_{\mathrm{AI}},p_{\mathrm{LD}})$ from $(0.7,0.3)$ to $(0.5,0.5)$; RA stage $(p_{\mathrm{AI}},p_{\mathrm{LD}},p_{\mathrm{RA}})$ from $(0.4,0.5,0.1)$ to $(0.3,0.3,0.4)$. We couple mode sampling with variable-rate training from \Cref{sec:Variable}: each iteration jointly samples a \emph{mode} (AI/LD/RA) and a \emph{rate} (its $\lambda$ and associated $(q_e,q_d,q_f,q_r,a)$) and optimizes only the selected path.

Staging temporal capability (LD$\rightarrow$RA), replaying earlier modes during optimization, and supervising the reliability gate stabilize training and prevent over-reliance on temporal cues. The gate learns to upweight temporal information under consistent motion and downweight it under contamination or scene cuts, preserving robustness without sacrificing the intra-baseline. After all these training stages, we perform fine-tuning on long sequences with randomly sampled modes.

\section{Experiment}
\subsection{Experimental settings}
\textbf{Training data:}  
MLIC training set~\cite{jiang2025mlic++} is utilized to pretrain the intra codec.
We use the training partition of the Vimeo-90k septuplet dataset~\cite{xue2019video} as the source of video training samples. BVI-AOM~\cite{nawala2024bvi}, LAVIB~\cite{stergiou2024lavib}, and processed original Vimeo videos~\cite{li2024neural} are used as the long sequences (64 frames) for error propagation-aware fine-tuning. Several sequences are removed to prevent data leakage.

\textbf{Training details.} Following CompressAI~\cite{begaint2020compressai}, we set $\lambda$ from $0.0009$ to $0.0483$ and interpolate 64 quality levels following~\cite{li2024neural}. EMA~\cite{moralesexponential} (decay=0.999) is enabled. BCE loss $\epsilon$ is set to 0.1, following common practice~\cite{dosovitskiy2020image}. Training was conducted on 8 NVIDIA H100 GPUs with the Adam.

\textbf{Testing data:}  
For testing, we evaluate our models on benchmark datasets widely used in the video compression literature: HEVC Class B~\cite{jvetj1010}; HEVC Class C~\cite{jvetj1010}; HEVC Class D~\cite{jvetj1010}; HEVC Class E~\cite{jvetj1010}; UVG~\cite{uvg2021}; MCL-JCV~\cite{wang2016mcl}.

\textbf{Test details.} For the traditional codec, we compare with VTM 18.0~\cite{bross2021overview} (QP22-47). For neural codecs, we evaluate open-source LICs for AI, including LALIC~\cite{feng2025linear}, DCAE~\cite{lu2025learned}, DCVC-RT AI~\cite{jia2025towards}, and HPCM~\cite{li2025learned}. For inter, we compare with open-source LVCs, including LD codecs such as DCVC-DC~\cite{li2023neural}, DCVC-FM~\cite{li2024neural}, DCVC-RT~\cite{jia2025towards}, DCMVC~\cite{tang2025neural}, and HyTIP~\cite{chen2025hytip}, as well as RA codecs such as DCVC-B~\cite{sheng2025bi} and BRHVC~\cite{liu2025neural}\footnote{BiECVC is excluded from comparison as no official implementation has been publicly released.}. We test 96 frames with an intra-period of -1 (IP-1) for a practical setting and an intra-period of 32 (IP32) for comparison with the RA setting. Please note that all the testing is in the RGB color space (BT.709) for fair comparison, so the results slightly differ from DCMVC~\cite{tang2025neural} and HyTIP~\cite{chen2025hytip}, which were originally tested on BT.601 RGBs. For the last GOP in the RA testing of DCVC-B, BRHVC, and Uni-LVC, we replaced the ending I-frame reference with the starting I-frame.

\textbf{VTM settings.}
\label{sec:vtm}
Following the suggestion from~\cite{li2023neural}, we first convert the original YUV420 videos to RGB content by the BT.709 matrix, and when utilizing the VTM to encode videos, we convert the content from RGB to YUV444 format. The internal color space is YUV444. We use the \texttt{encoder\_lowdelay\_vtm.cfg} configuration file for the LD anchor and \texttt{encoder\_randomaccess\_vtm.cfg} configuration file for RA anchor. 
QP values of 22,27,32,37,42,47 are tested. After encoding and decoding, the reconstructed YUV444 sequences are converted back to the RGB color space for the PSNR calculation.

\subsection{Quantitative results}
\begin{figure*}[htbp]
    \centering
    \subfloat[HEVC-B\label{subfig:hecv-b-intra}]{
        \includegraphics[width=0.3\linewidth]{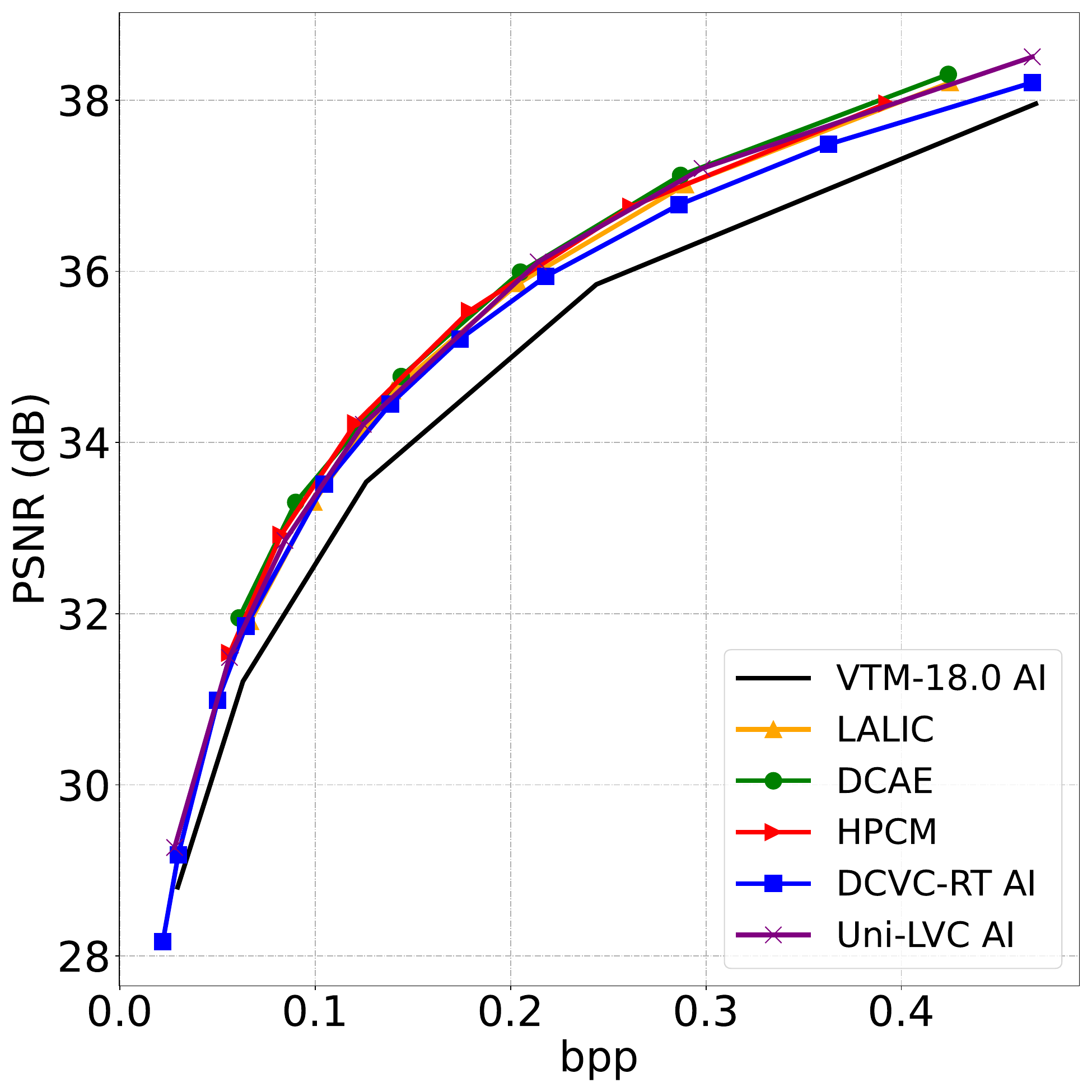}
    }
    \hfill
    \subfloat[HEVC-C\label{subfig:hecv-c-intra}]{
        \includegraphics[width=0.3\linewidth]{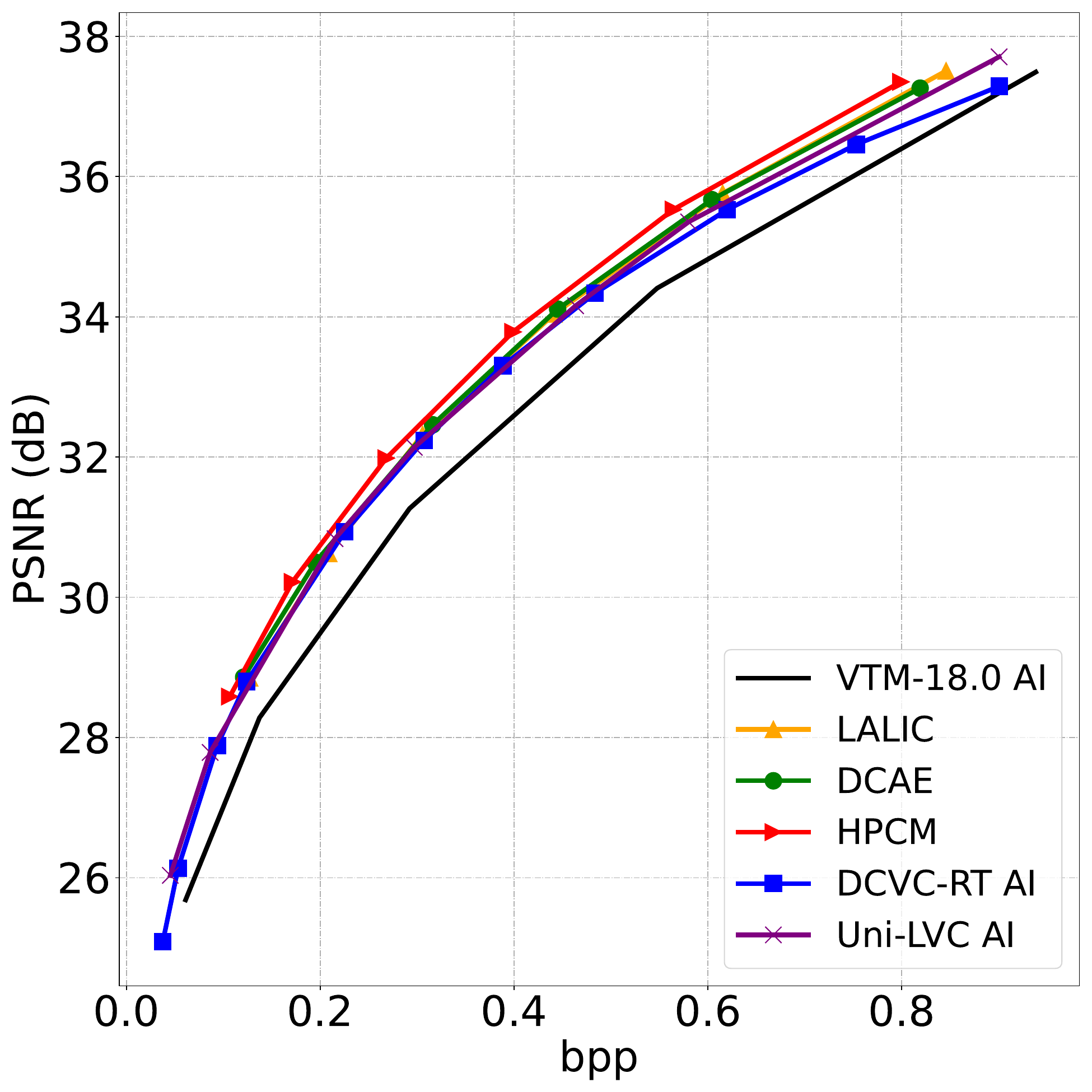}
    }
    \hfill
    \subfloat[HEVC-D\label{subfig:hecv-d-intra}]{
        \includegraphics[width=0.3\linewidth]{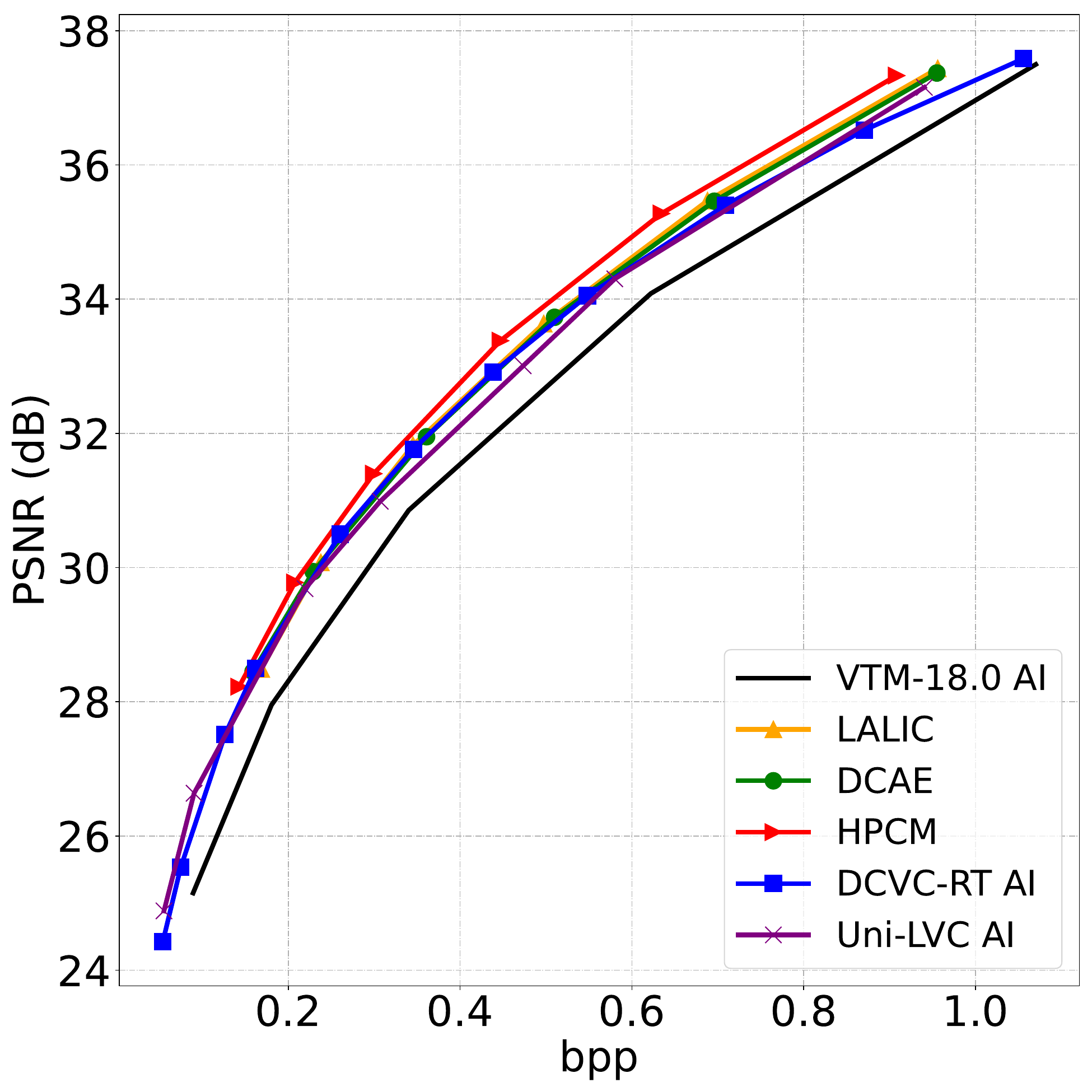}
    }
    \hfill
    \subfloat[HEVC-E\label{subfig:hecv-e-intra}]{
        \includegraphics[width=0.3\linewidth]{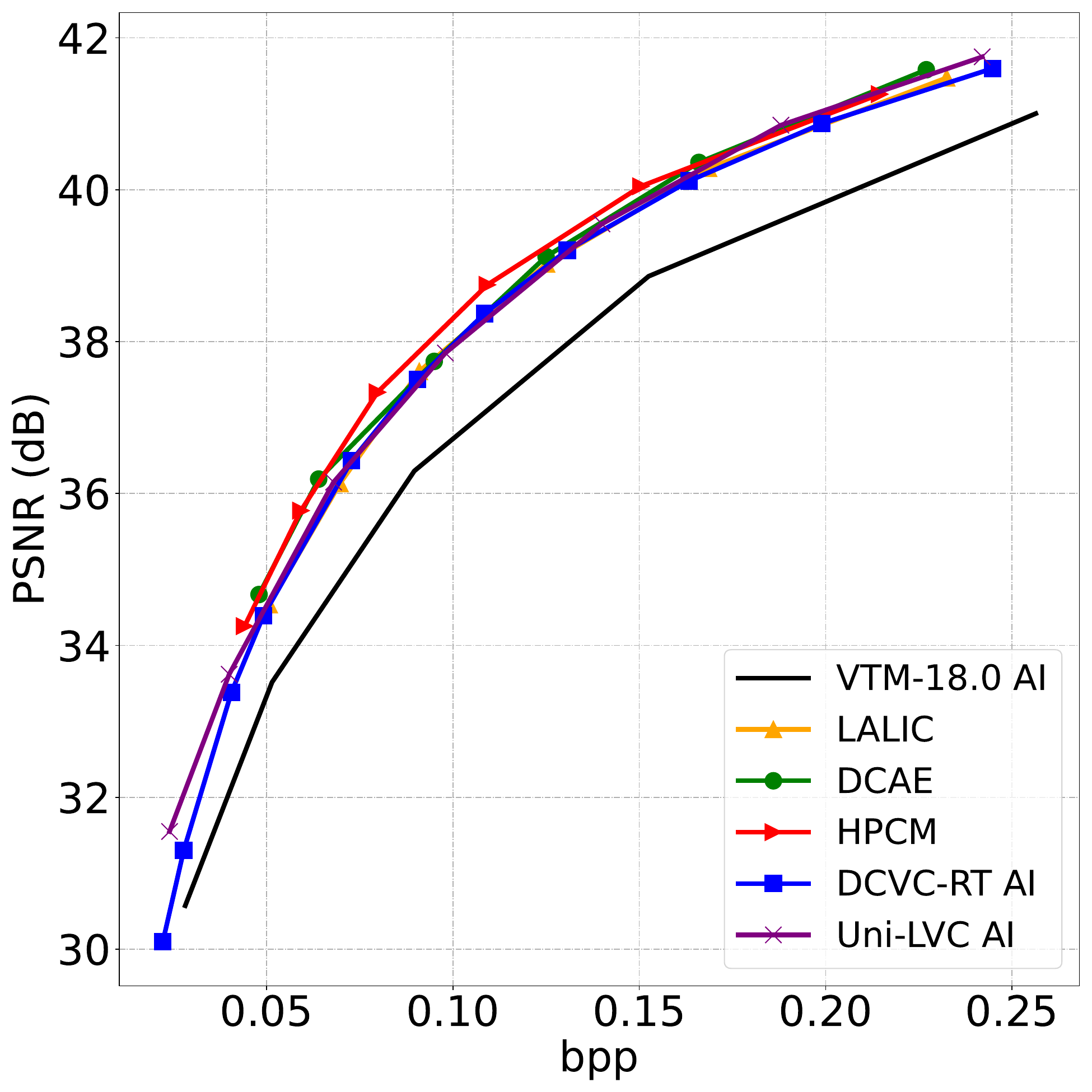}
    }
    \hfill
    \subfloat[UVG\label{subfig:uvg-intra}]{
        \includegraphics[width=0.3\linewidth]{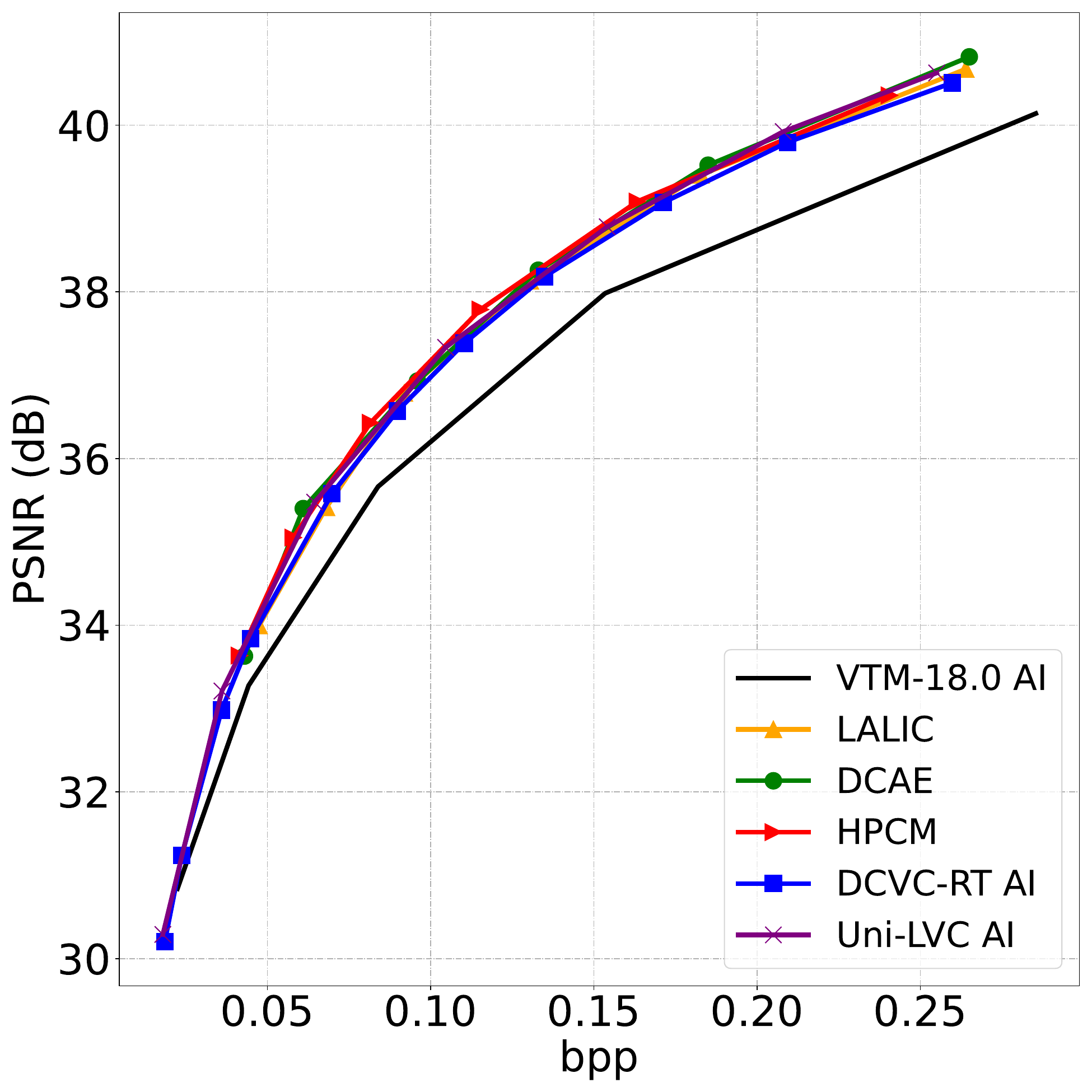}
    }
    \hfill
    \subfloat[MCL-JCV\label{subfig:mcl-jcv-intra}]{
        \includegraphics[width=0.3\linewidth]{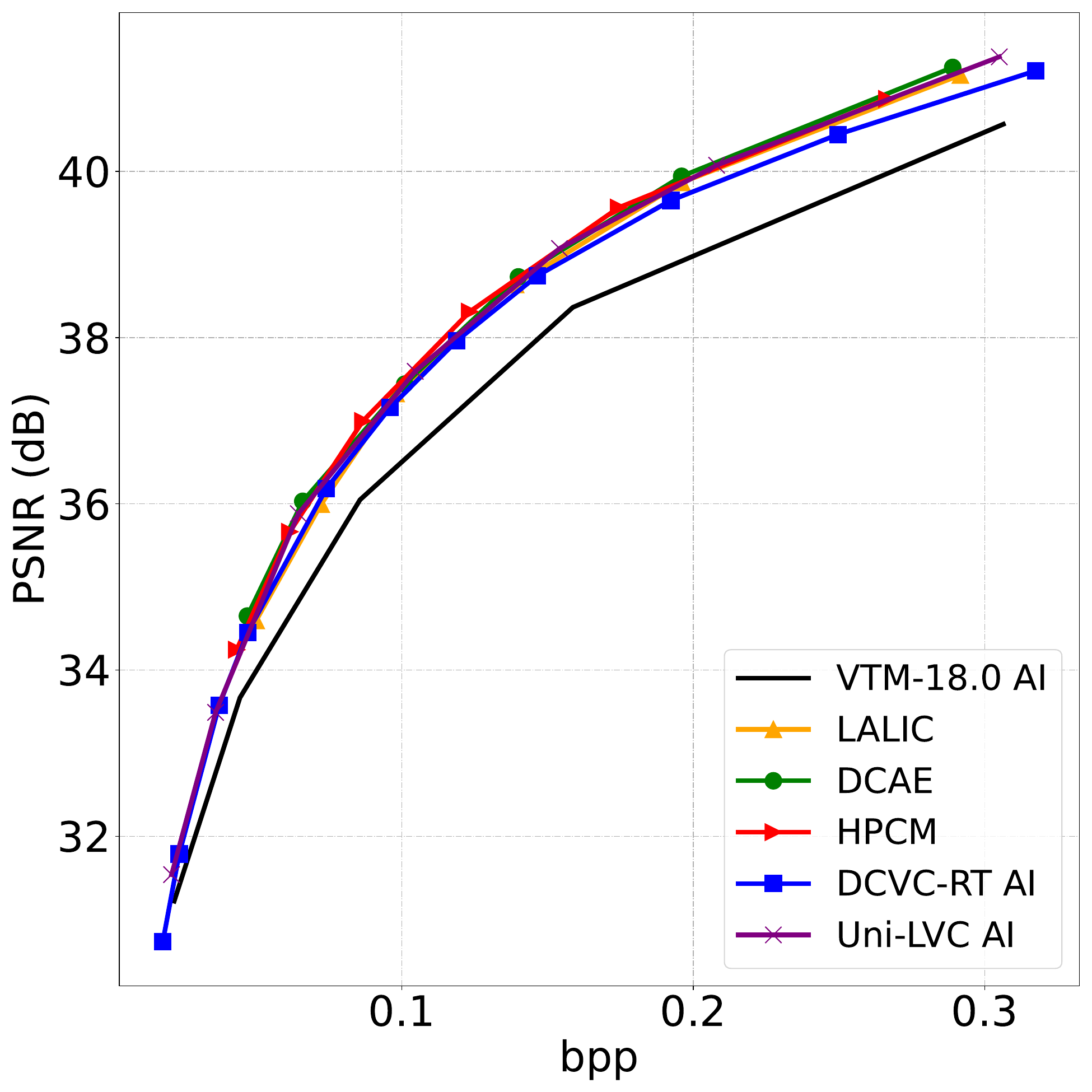}
    }
    \caption{\textbf{Rate-distortion (R-D) curves for various methods for intra coding.} \textit{Please zoom in for more details.}}
    \label{fig:rd_fig}
\end{figure*}

\begin{figure*}[htbp]
    \centering
    \subfloat[HEVC-B\label{subfig:hecv-b-inter}]{
        \includegraphics[width=0.3\linewidth]{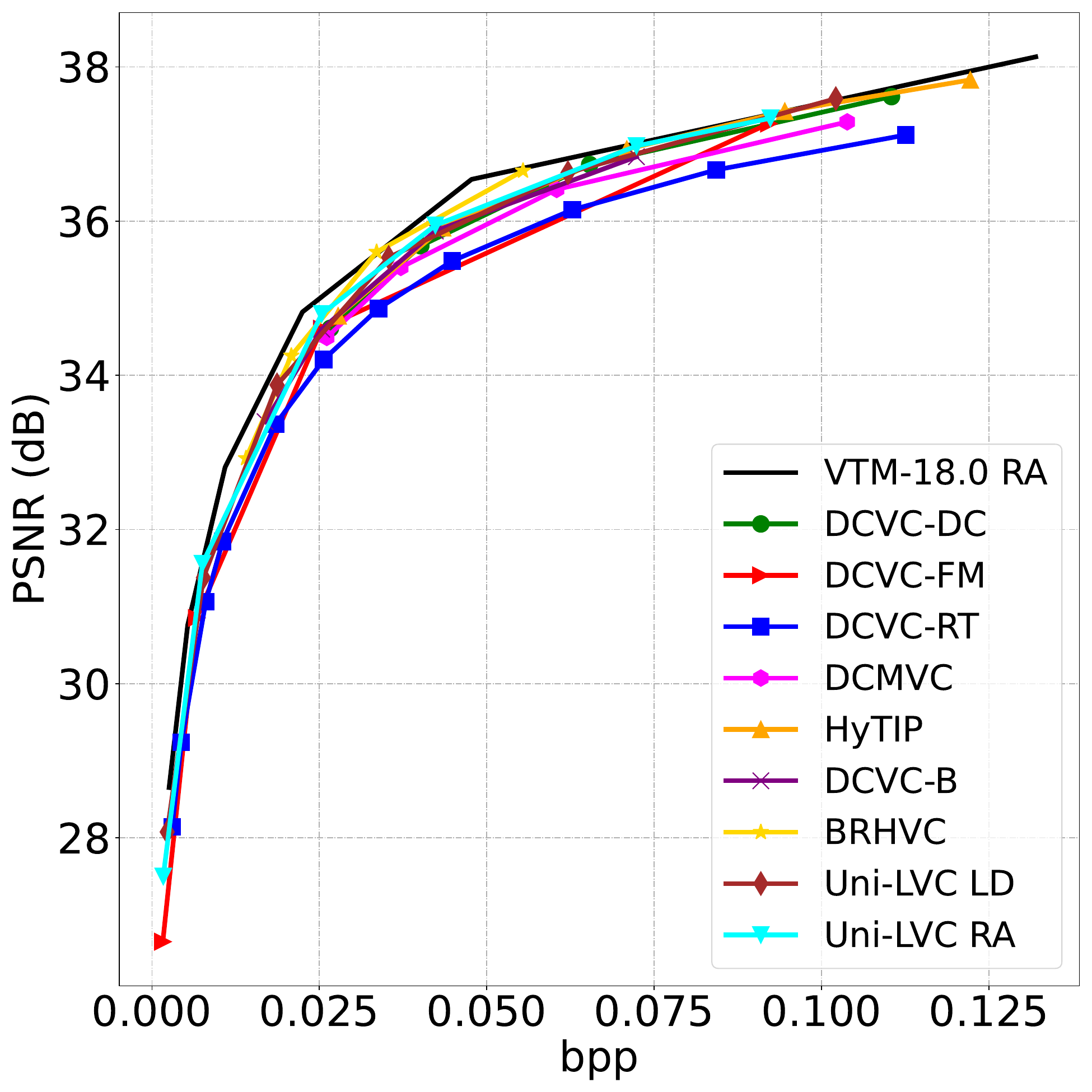}
    }
    \hfill
    \subfloat[HEVC-C\label{subfig:hecv-c-inter}]{
        \includegraphics[width=0.3\linewidth]{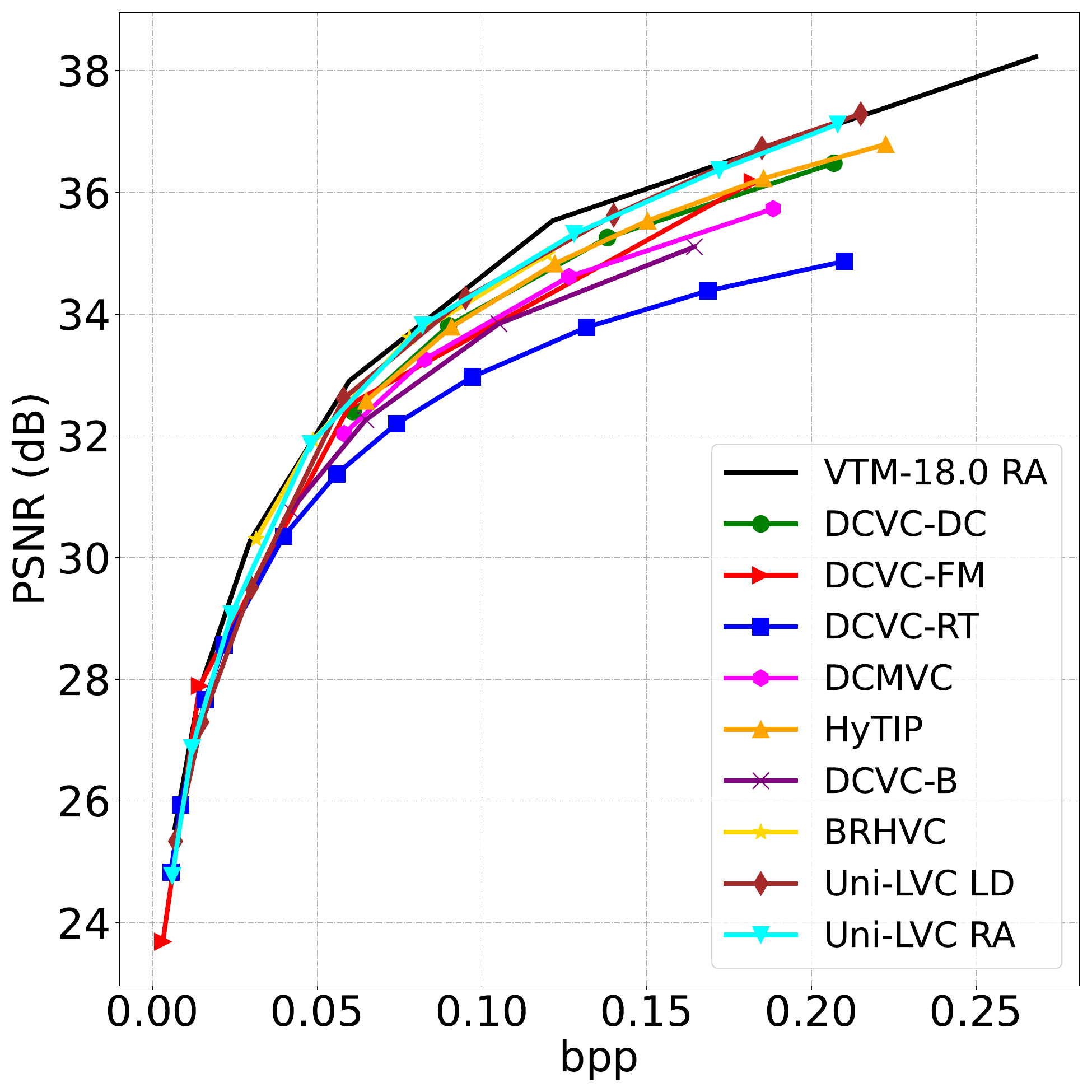}
    }
    \hfill
    \subfloat[HEVC-D\label{subfig:hecv-d-inter}]{
        \includegraphics[width=0.3\linewidth]{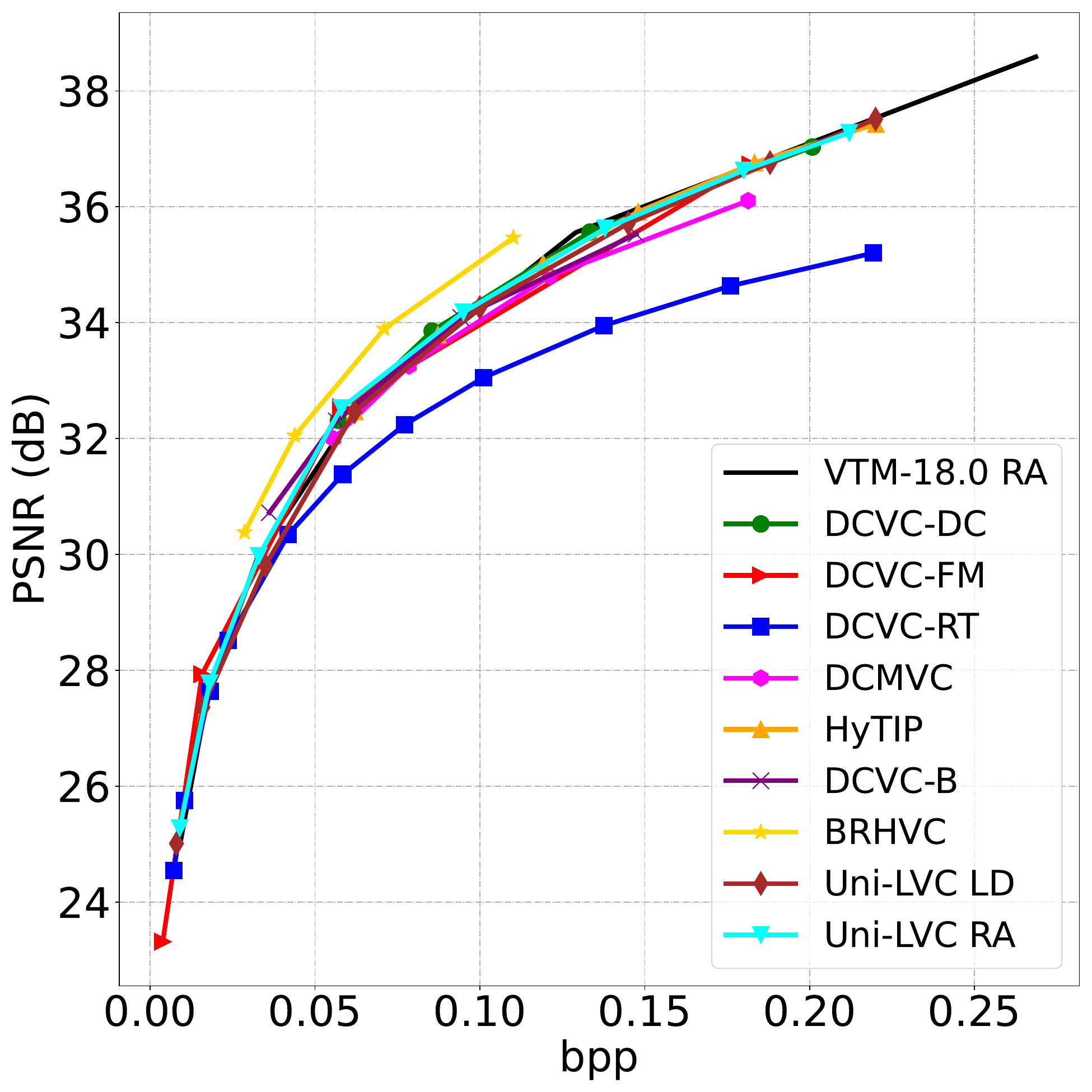}
    }
    \hfill
    \subfloat[HEVC-E\label{subfig:hecv-e-inter}]{
        \includegraphics[width=0.3\linewidth]{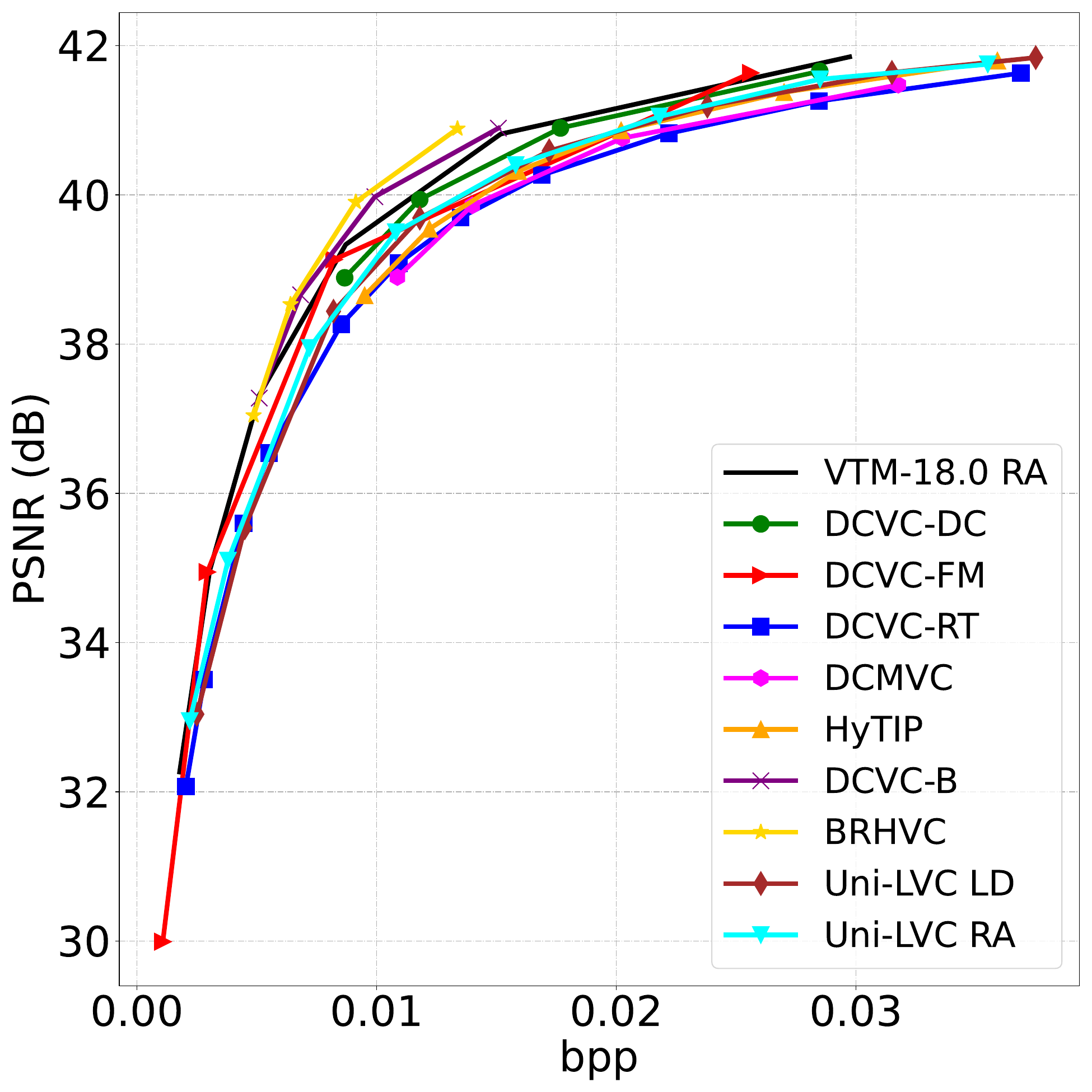}
    }
    \hfill
    \subfloat[UVG\label{subfig:uvg-inter}]{
        \includegraphics[width=0.3\linewidth]{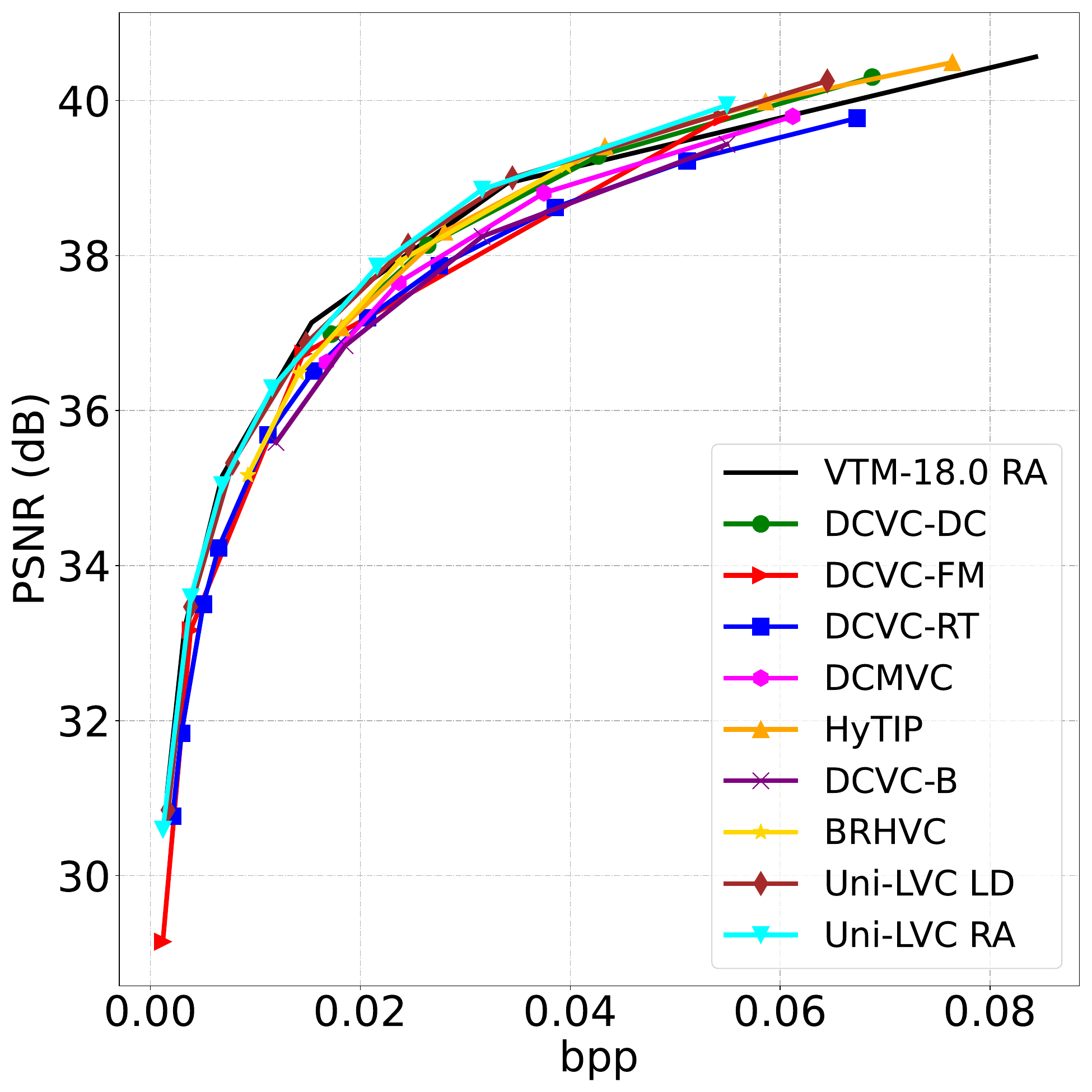}
    }
    \hfill
    \subfloat[MCL-JCV\label{subfig:mcl-jcv-inter}]{
        \includegraphics[width=0.3\linewidth]{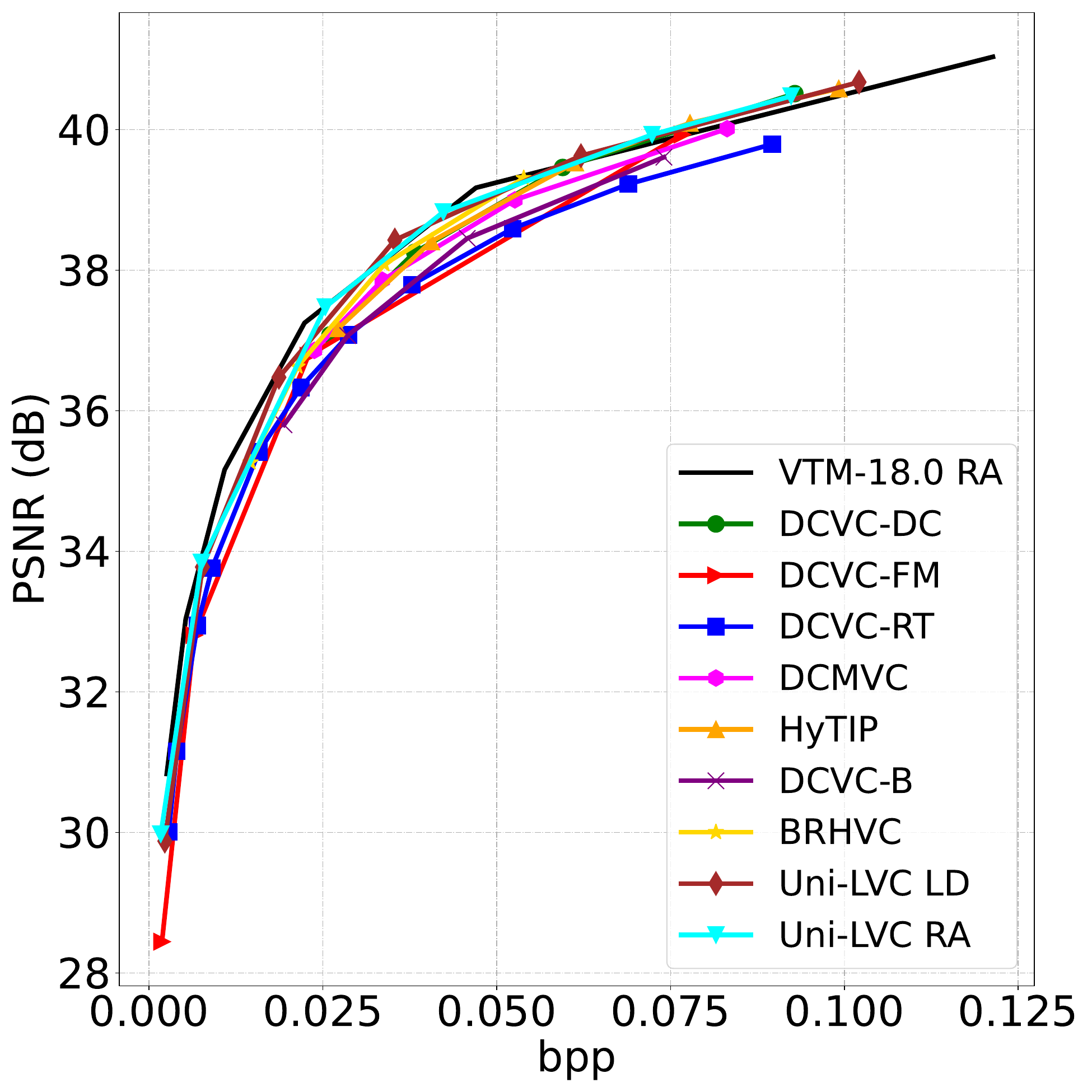}
    }
    \caption{\textbf{Rate-distortion (R-D) curves for various methods for inter coding, intra period of 32.} \textit{Please zoom in for more details.}}
    \label{fig:rd_fig_inter}
\end{figure*}

We report R–D results and computational cost in \Cref{tab:bdrate}. R-D curves are shown in ~\Cref{fig:rd_fig} and ~\Cref{fig:rd_fig_inter}.

\begin{table*}[htbp]
    \centering
    \small
    \caption{Computational Complexity and BD-Rate Compared to Existing Methods. Best result is \underline{underlined}. \textbf{Bold} denotes our method.}
    \resizebox{\textwidth}{!}{%
    \begin{threeparttable}
    \begin{tabular}{c|c|c|c|c|c|c|c|c|c|c|c}
    \hline\hline
    \multirow{2}{*}{Method} & \multirow{2}{*}{Type} & \multirow{2}{*}{Total Params. $\downarrow$} & \multicolumn{2}{c|}{Latency $\downarrow$} & \multicolumn{7}{c}{BD-Rate (\%) w.r.t. VTM 18.0 $\downarrow$} \\
    \cline{4-12}
     &  &  & Encoding & Decoding & \makecell{HEVC\\Class B} & \makecell{HEVC\\Class C} & \makecell{HEVC\\Class D} & \makecell{HEVC\\Class E} & UVG & MCL-JCV & Average \\
    \hline
    VTM-18.0 & AI & - & - & - & 0\% & 0\% & 0\% & 0\% & 0\% & 0\% & 0\% \\
    LALIC~\cite{feng2025linear} & AI & 396.8M & 0.491s & 0.165s & -16.57\% & -15.06\% & -14.54\% & -20.60\% & -16.92\% & -17.46\% & -16.86\% \\
    DCAE~\cite{lu2025learned} & AI & 716.4M & 0.346s & 0.353s & \underline{-20.76\%} & -16.17\% & -14.26\% & -23.71\% & {-19.29\%} & {-20.62\%} & -19.14\% \\
    HPCM~\cite{li2025learned} & AI & 538.3M & 0.232s & 0.212s & -19.88\% & \underline{-20.24\%} & \underline{-19.56\%} & \underline{-26.10\%} & \underline{-19.89\%} & \underline{-20.76\%} & \underline{-21.07\%} \\
    DCVC-RT AI~\cite{jia2025towards} & AI & \underline{45.7M} & \underline{0.023s} & \underline{0.018s} & -13.85\% & -15.91\% & -15.67\% & -19.45\% & -13.84\% & -14.77\% & -15.58\% \\
    \textbf{Uni-LVC} & \textbf{AI} & \textbf{50.5M} & \textbf{0.071s} & \textbf{0.062s} & \textbf{-18.41\%} & \textbf{-18.30\%} & \textbf{-16.50\%} & \textbf{-22.99\%} & \textbf{-17.93\%} & \textbf{-18.43\%} & \textbf{-18.76\%} \\\hline
    VTM-18.0 & LD IP-1 & - & - & - & 0\% & 0\% & 0\% & 0\% & 0\% & 0\% & 0\% \\
    DCVC-DC~\cite{li2023neural} & LD IP-1 & 50.8M & 0.442s & 0.358s & -11.90\% & -1.32\% & -23.47\% & -15.09\% & -17.79\% & -11.58\% & -13.53\% \\
    DCVC-FM~\cite{li2024neural} & LD IP-1 & \underline{44.9M} & 0.459s & 0.382s & -16.37\% & \underline{-20.72\%} & \underline{-38.54\%} & \underline{-27.59\%} & -22.34\% & -9.56\% & \underline{-22.52\%} \\
    DCVC-RT~\cite{jia2025towards} & LD IP-1 & 66.3M & \underline{0.022s} & \underline{0.022s} & -10.74\% & -5.08\% & -20.23\% & -12.34\% & -17.56\% & -9.94\% & -12.65\% \\
    DCMVC~\cite{tang2025neural} & LD IP-1 & 52.0M & 0.707s & 0.626s & -12.65\% & -0.62\% & -22.28\% & -3.03\% & -13.91\% & -12.30\% & -10.80\% \\
    HyTIP~\cite{chen2025hytip} & LD IP-1 & 47.6M & 0.599s & 0.504s & -15.92\% & -5.99\% & -25.28\% & 1.57\% & -23.74\% & -15.99\% & -14.75\% \\
    \textbf{Uni-LVC} & \textbf{LD IP-1} & \textbf{65.1M} & \textbf{0.073s} & \textbf{0.065s} & \textbf{\underline{-20.99\%}} & \textbf{-9.74\%} & \textbf{-29.47\%} & \textbf{-6.78\%} & \textbf{\underline{-25.51\%}} & \textbf{\underline{-19.39\%}} & \textbf{-18.65\%} \\\hline
    VTM-18.0 & RA IP32 & - & - & - & 0\% & 0\% & 0\% & 0\% & 0\% & 0\% & 0\% \\
    DCVC-DC~\cite{li2023neural} & LD IP32 & 50.8M & 0.442s & 0.358s & 23.27\% & 20.70\% & 1.21\% & 9.22\% & 7.54\% & 13.25\% & 12.53\% \\
    DCVC-FM~\cite{li2024neural} & LD IP32 & \underline{44.9M} & 0.459s & 0.382s & 22.45\% & \underline{6.47\%} & \underline{-11.59\%} & \underline{-1.33\%} & 14.18\% & 25.59\% & \underline{9.30\%} \\
    DCVC-RT~\cite{jia2025towards} & LD IP32 & 66.3M & \underline{0.022s} & \underline{0.022s} & 43.13\% & 36.33\% & 19.57\% & 30.74\% & 34.75\% & 36.94\% & 33.58\% \\
    DCMVC~\cite{tang2025neural} & LD IP32 & 52.0M & 0.707s & 0.626s & 32.45\% & 30.31\% & 9.58\% & 34.33\% & 20.74\% & 19.62\% & 24.51\% \\
    HyTIP~\cite{chen2025hytip} & LD IP32 & 47.6M & 0.599s & 0.504s & 19.69\% & 21.98\% & 4.12\% & 28.05\% & 5.36\% & 13.79\% & 15.50\% \\
    \textbf{Uni-LVC} & \textbf{LD IP32} & \textbf{65.1M} & \textbf{0.073s} & \textbf{0.065s} & \textbf{\underline{16.71\%}} & \textbf{15.17\%} & \textbf{2.98\%} & \textbf{25.21\%} & \textbf{\underline{5.21\%}} & \textbf{\underline{10.47\%}} & \textbf{12.63\%} \\\hdashline
    DCVC-B~\cite{sheng2025bi} & RA IP32 & \underline{55.4M} & 0.729s & 0.587s & 24.11\% & 33.12\% & 0.14\% & -3.96\% & 33.61\% & 34.64\% & 20.28\% \\
    BRHVC~\cite{liu2025neural} & RA IP32 & {60.4M} & 1.115s & 0.779s & 14.20\% & \underline{7.43\%} & \underline{-18.80\%} & \underline{-7.67\%} & 16.83\% & 17.26\% & \underline{4.88\%} \\
    \textbf{Uni-LVC} & \textbf{RA IP32} & \textbf{65.1M} & \textbf{\underline{0.075s}} & \textbf{\underline{0.065s}} & \textbf{\underline{10.31\%}} & \textbf{10.61\%} & \textbf{-1.11\%} & \textbf{19.85\%} & \textbf{\underline{0.98\%}} & \textbf{\underline{5.31\%}} & \textbf{7.66\%} \\
    \hline\hline
    \end{tabular}%
    \begin{tablenotes}\small
      \item \textbf{Test conditions:} Nvidia 4090 GPU. The enc./dec. time is averaged over 1080p test sequences, including entropy enc./dec. time.
      \item For the AI mode, VTM-18.0 AI is used as the anchor. For LD modes with an intra-period of -1, VTM-18.0 LD is used as the anchor. For the LD and RA modes with an intra-period of 32, VTM-18.0 RA is used as the anchor.
    \end{tablenotes}
    \end{threeparttable}    }
    \label{tab:bdrate}
\end{table*}

\textbf{Evaluation criteria.}
(1) \emph{BD-Rate}: the average percentage bitrate difference at matched quality between two R–D curves (lower is better). Unless otherwise noted, we compute BD-Rate with respect to VTM-18.0 (anchor) over PSNR~\cite{BDrate}. 
(2) \emph{Latency}: per-frame wall-clock time (seconds) for encoding and decoding, including arithmetic coding/decoding, measured on an NVIDIA~4090 (lower is better).
(3) \emph{Model size}: total trainable parameters (millions; lower is better).  
For {all-intra codecs} (image codecs), we report the \emph{total parameters required to cover the full bitrate–quality operating range}. {Fixed-rate methods} (e.g., LALIC) train separate models per target bitrate. Their model size is therefore the \emph{sum} of all models needed to span the reported range (e.g., 6 models $\Rightarrow$ total parameters = $6\times$ per-model size). {Variable-rate methods} (e.g., Uni-LVC, DCVC-RT AI) use a \emph{single} model with internal rate-control mechanisms; thus the parameter count is reported once.
For {inter codecs} (video codecs), both the intra and the inter modules are required at inference. We therefore report the \emph{combined} parameter count:
\(
\text{Model size (inter)} = \text{intra parameters} + \text{inter parameters}.
\)
All neural codecs use identical hardware settings for fair comparison.

\subsubsection{All-intra (AI)}
\label{sec:ai_results}
\textbf{R-D efficiency.} Uni-LVC (AI) attains an average BD-Rate of $-18.76\%$, improving over DCVC-RT AI $(-15.58\%)$ by $-3.18\%$ and approaching the accuracy of much larger models such as HPCM $(-21.07\%)$ and DCAE $(-19.14\%)$.

\textbf{Runtime and model size.} With $50.5$M parameters, Uni-LVC (AI) is ${\sim}10.7\times$ and ${\sim}14.2\times$ smaller than HPCM (538.3M) and DCAE (716.4M), respectively, while remaining within ${\leq}2.31\%$ BD-Rate of both. Despite being slightly larger than DCVC-RT AI (45.7M), Uni-LVC improves BD-Rate by $-3.18\%$. In latency, Uni-LVC encodes $3.3$--$6.9\times$ faster and decodes $2.7$--$5.7\times$ faster than LALIC/DCAE/HPCM, while running ${\sim}3\times$ slower than DCVC-RT AI due to our stronger entropy modeling, quantization, and the ultra-optimized DCVC-RT implementation~\footnote{By incorporating efficient implementations from DCVC-RT, e.g., low precision inference, efficient CUDA kernels, we believe the inference speed of Uni-LVC could be further improved}.

\subsubsection{Low-delay (LD)}
\label{sec:ld_results}
\textbf{R-D efficiency.} Uni-LVC (LD) achieves an average BD-Rate of $-18.65\%$ (IP$=-1$) relative to VTM-18.0 (LD), outperforming HyTIP $(-14.75\%)$ by $3.90\%$, DCVC-DC $(-13.53\%)$ by $5.12\%$, DCVC-RT $(-12.65\%)$ by $6.00\%$, and DCMVC $(-10.80\%)$ by $7.85\%$, while approaching DCVC-FM $(-22.52\%)$. Notably, Uni-LVC demonstrates particularly strong performance on 1080p sequences under both IP settings. Under IP$=-1$, Uni-LVC achieves the best BD-Rate on HEVC Class~B ($-20.99\%$), UVG ($-25.51\%$), and MCL-JCV ($-19.39\%$), outperforming all competing methods, including DCVC-FM on these datasets. The same trend is observed under IP$=32$, where Uni-LVC leads on HEVC Class~B ($16.71\%$), UVG ($5.21\%$), and MCL-JCV ($10.47\%$). On lower-resolution sequences (HEVC Class~C, D, and E), DCVC-FM retains an advantage under both settings (e.g., $-38.54\%$ vs.\ $-29.47\%$ on Class~D under IP$=-1$; $-11.59\%$ vs.\ $2.98\%$ under IP$=32$), suggesting that the optical flow motion is useful for lower-resolution content~\cite{bilican2025content} or static content, whereas Uni-LVC generalizes more effectively to high-resolution video. Under the IP$=32$ setting against the VTM-18.0 RA anchor, Uni-LVC (LD) achieves an average of $12.63\%$, outperforming HyTIP ($15.50\%$) by $2.87\%$, DCMVC ($24.51\%$) by $11.88\%$, and DCVC-RT ($33.58\%$) by $20.95\%$, with results comparable to DCVC-DC ($12.53\%$) and approaching DCVC-FM ($9.30\%$).

\textbf{Runtime and model size.} Uni-LVC (LD) runs at $0.073$s\,/\,$0.065$s (enc/dec) with $65.1$M parameters. This is ${\sim}6.1\times$ faster to encode than DCVC-DC and ${\sim}6.3\times$ faster than DCVC-FM (decode: ${\sim}5.5$--$5.9\times$), with comparable parameters. DCVC-RT offers lower latency but requires $20.95\%$ more bitrate on average under the IP$=32$ RA comparison.\footnote{DCVC-RT's poor performance mainly comes from the higher BPP range ($>$0.05 BPP) due to its simple efficiency model design. On the lower BPP range, it is comparable to DCVC-FM.}

\subsubsection{Random-access (RA)}
\label{sec:ra_results}
\textbf{R-D efficiency.} Uni-LVC (RA) reaches an average BD-Rate of $7.66\%$, improving over DCVC-B ($20.28\%$) by $12.62\%$. Compared with BRHVC ($4.88\%$), Uni-LVC trails by $2.78\%$ on average; however, Uni-LVC demonstrates stronger performance on 1080p sequences. On HEVC Class~B, UVG, and MCL-JCV, Uni-LVC achieves $10.31\%$, $0.98\%$, and $5.31\%$, respectively, outperforming DCVC-B ($24.11\%$, $33.61\%$, $34.64\%$) by large margins. BRHVC holds an advantage on lower-resolution and static content sequences, achieving the best BD-Rate on HEVC Class~C ($7.43\%$), Class~D ($-18.80\%$), and Class~E ($-7.67\%$), which is similar in LD settings. On high-resolution 1080p sequences, Uni-LVC consistently leads all RA methods.

\textbf{Runtime and model size.} Uni-LVC (RA) maintains nearly the same latency as LD (\(0.075\)s\,/\,\(0.065\)s) with the same \(65.1\)M parameters, while DCVC-B is \(\sim 9.7\times\) slower in encoding and \(\sim 9.0\times\) slower in decoding. BRHVC incurs even greater latency at $1.115$s\,/\,$0.779$s (enc/dec), making Uni-LVC ${\sim}14.9\times$ faster to encode and ${\sim}12.0\times$ faster to decode, at a cost of $2.78\%$ average BD-Rate.

Overall, our Uni-LVC utilizes a \emph{single} model that supports AI, LD, and RA, and consists of one intra backbone plus a temporal extension. In contrast, prior DCVC approaches maintain separate models for AI/LD/RA. Uni-LVC (AI) approaches the performance of large-scale intra codecs with ${\sim}10\times$ fewer parameters; Uni-LVC (LD) delivers competitive BD-Rate among LD codecs with significantly lower latency than DCVC-FM and DCVC-DC while outperforming all remaining LD methods; and Uni-LVC (RA) surpasses DCVC-B by $12.62\%$ with ${\sim}10\times$ lower latency, and achieves competitive average BD-Rate within $2.78\%$ of BRHVC while running ${\sim}14.9\times$ faster.

\subsection{Ablation study}
\label{sec:abstudy}
All ablations are conducted on the HEVC Class~B sequences under the same test conditions as the main experiment. For each ablation step, we manually control the parameters/complexity around the same level to make the comparison fair.
Unless otherwise specified, we report BD-Rate (\%) against the VTM 18.0, AI for intra and RA IP32 for inter (lower is better). 
We analyze the effect of (i) the intra-codec components, (ii) the temporal adaptation and reliability-aware classifier, and (iii) the training strategy.

\subsubsection{Intra codec components}
We first ablate the intra codec by progressively adding the proposed components on top of a re-implementation of DCVC-RT in the all-intra (AI) configuration. Starting from the baseline (B0), we successively add: (1) the enhanced DC blocks and cross-attention, (2) the HPCM model, and (3) the lattice vector quantizer (LVQ) and density scaling. The full intra configuration of the proposed unified codec then corresponds to B3. Table~\ref{tab:ablation_intra} summarizes the BD-Rate results on HEVC Class~B in the all-intra (AI) setting. 

\begin{table}[htbp]
  \centering
  \caption{Ablation on intra codec components on HEVC Class~B in the AI configuration. 
  Anchor is VTM 18.0 AI (lower is better).}
  \label{tab:ablation_intra}
  \begin{tabular}{c|c}
    \hline \hline
    Variant & BD-Rate (AI, Class B) \\
    \hline 
    B0: DCVC-RT (re-impl)          & -12.35\%\\ 
    B1: B0 + enhanced blocks              & -14.03\% \\ 
    B2: B1 + HPCM                             & -16.05\% \\ 
    B3: B2 + LVQ (ours final)                      & -18.41\% \\ 
     \hline  \hline 
  \end{tabular}
\end{table}

B0 corresponds to our re-implementation of DCVC-RT in the AI setting and serves as the baseline.  
Adding the enhanced DC blocks and the cross-attention module (B1) already yields a noticeable gain of 1.68\% in BD-Rate (from $-12.35\%$ to $-14.03\%$), indicating that the improved spatio-channel mixing and the conditional design are beneficial even in the all-intra regime.  
Introducing the hierarchical progressive context model (HPCM) in B2 further reduces BD-Rate to $-16.05\%$, contributing another $\sim 2\%$ improvement through stronger entropy modeling.  
Finally, equipping the codec with lattice vector quantization and density scaling (B3) brings the BD-Rate to $-18.41\%$, a total gain of roughly 6\% over the baseline B0.  
These improvements confirm that each intra component contributes meaningfully to rate-distortion performance, and that the full intra configuration of Uni-LVC (B3) is substantially stronger than the DCVC-RT baseline while remaining efficient.

\subsubsection{Temporal adaptation and reliability-aware classifier}
We next study the temporal adaptation modules and the reliability-aware classifier in the inter. 
On top of the intra model, we add the temporal buffer and temporal adaptation with local deformable 
neighborhood cross-attention (DN-CA) and global polarity-aware linear cross-attention (PAL-CA), as well as the classifier that decides the reliability of temporal features and scales the feature correspondingly.

We consider the following variants:
\begin{itemize}
    \item I0: directly concatenate temporal features to the intra component and squeeze channels;
    \item I1: temporal path without classifier, using a Swin Transformer-based cross-attention~\cite{chen2021crossvit} (buffer + Swin-CA, always using temporal features, recurrent update);
    \item I2: replace Swin-CA with proposed DN-CA;
    \item I3: add proposed PAL-CA to I2;
    \item I4: add the reliability-aware classifier (our final);
\end{itemize}

Table~\ref{tab:ablation_inter} reports BD-Rate results on HEVC Class~B under the low-delay (LD), IP 32 
configuration. 

\begin{table}[htbp]
  \centering
  \caption{Ablation on temporal adaptation and the reliability-aware classifier on HEVC Class~B. Anchor is VTM 18.0 RA (lower is better).}
  \label{tab:ablation_inter}
  \resizebox{\linewidth}{!}{
  \begin{tabular}{c|c}
    \hline \hline
    Variant & BD-Rate (LD, Class B) \\
    \hline 
    I0: direct concatenate temporal features        & 26.71\%\\
    I1: temporal path (Swin-CA), no classifier        & 25.97\%\\
    I2: replace Swin-CA with DN-CA                   & 22.93\%\\
    I3: add PAL-CA to I2                             & 18.63\% \\
    I4: add the reliability-aware classifier (ours final)  & 16.71\% \\
     \hline  \hline 
  \end{tabular}
  }
\end{table}

Starting from I0, where temporal features are naively integrated by direct concatenation and channel squeezing, the BD-Rate is $26.71\%$ relative to VTM-18.0 in the LD IP32 setting. Despite its simplicity, this already provides a baseline for temporal exploitation, but the lack of structured attention limits the model's ability to selectively leverage useful temporal cues. Replacing the naive concatenation with a Swin Transformer-based cross-attention module (Swin-CA) in I1 yields an improvement to $25.97\%$, confirming that cross-attention is 
preferable to simple feature fusion, though the gain is limited by the generic windowed attention design. Replacing Swin-CA with the proposed deformable neighborhood cross-attention (DN-CA) in I2 
improves BD-Rate to $22.93\%$, demonstrating that motion-aware local attention is more effective 
than generic windowed attention for temporal prediction. Adding the global polarity-aware linear cross-attention (PAL-CA) on top of DN-CA in I3 further 
reduces BD-Rate to $18.63\%$, indicating that combining deformable local matching with efficient 
global temporal modeling significantly enhances inter coding efficiency.
Finally, enabling the reliability-aware classifier in I4 yields an additional gain, reaching $16.71\%$ BD-Rate and corresponding to the final Uni-LVC (LD) configuration. The gap between I3 and I4 ($18.63\%\rightarrow16.71\%$) confirms that explicitly suppressing unreliable temporal cues, rather than always trusting the buffer, is important for robustness and overall rate-distortion performance.

\subsubsection{Training strategy}
We further investigate the effect of the training strategy for our codec. 
We compare three schemes using the same architecture but different curricula:

\begin{itemize}
  \item T0: single-stage joint training of AI/LD/RA from scratch;
  \item T1: sequential training without replay (AI $\rightarrow$ LD $\rightarrow$ RA, without preserving earlier modes);
  \item T2: staged training with mode sampling and knowledge replay (our proposed strategy).
\end{itemize}

All models are evaluated on HEVC Class~B for AI, LD, and RA configurations. 
Table~\ref{tab:ablation_training} reports the corresponding BD-Rate results; AI results anchor is VTM-18.0 AI, and LD/RA results anchor is VTM-18.0 RA.

\begin{table}[htbp]
  \centering
  \caption{Ablation on the training strategy on HEVC Class~B. 
  Numbers are BD-Rate (\%) versus the VTM reference for AI, LD, and RA configurations (lower is better).}
  \label{tab:ablation_training}
  \resizebox{\linewidth}{!}{
  \begin{tabular}{c|c|c|c}
    \hline \hline
    Scheme & AI & LD & RA \\
    \hline 
    T0: single-stage joint training        & -10.19\% & 27.9\% & 20.1\%  \\
    T1: sequential w/o replay              & 40.51\%  & 60.56\% & 9.01\% \\
    T2: staged + replay (ours final)       & -18.41\% & 16.71\% & 10.31\% \\
     \hline  \hline 
  \end{tabular}
  }
\end{table}

With single-stage joint training (T0), the unified model achieves moderate performance across all modes, but falls short of the best our proposed strategy: AI BD-Rate degrades to $-10.19\%$ (vs.\ $-18.41\%$ for the dedicated intra model), while LD and RA remain at $27.9\%$ and $20.1\%$, respectively.  
Sequential training without replay (T1) strongly biases optimization toward the last-trained RA mode: RA BD-Rate improves to $9.01\%$, but AI and LD suffer from severe catastrophic forgetting, with BD-Rates of $40.51\%$ and $60.56\%$, respectively, i.e., substantially worse than the VTM reference.  
Our staged training with mode sampling and knowledge replay (T2) achieves the best overall trade-off: AI attains $-18.41\%$, LD reaches $16.71\%$, and RA obtains $10.31\%$.  
Compared to T0, T2 improves all three modes simultaneously, and compared to T1, it preserves strong RA performance while avoiding the catastrophic degradation in AI and LD.  
These results validate the necessity of the proposed curriculum and replay strategy for maintaining balanced performance across all coding modes in a single unified model.

\subsubsection{Reliability-aware classifier and temporal path design}
\label{sec:abstudy_classifier}
We further ablate the design choices for the reliability-aware classifier and the temporal path, 
all evaluated on HEVC Class~B under the LD IP32 with VTM-18.0 RA as the anchor.

\paragraph{Gating design.}
Our classifier predicts a single scalar gate $\alpha_t \in [0,1]$ per frame, conditioned on both 
the current input frame $x_t$ and the temporal feature $f_{t-1}^*$. The scalar design is 
intentional: it keeps side information overhead negligible (one 16-bit value per frame) while 
still allowing the model to suppress unreliable temporal cues globally. As an alternative, one 
could predict a spatial gate map, but this either requires transmitting the map as side 
information (encoder-side) or forfeiting access to the original frame (decoder-side). 
Table~\ref{tab:ablation_gate} summarizes the comparison.

\begin{table}[htbp]
  \centering
  \caption{Ablation on gating design on HEVC Class~B, LD IP32. 
  Anchor is VTM 18.0 RA (lower is better).}
  \label{tab:ablation_gate}
  \resizebox{\linewidth}{!}{
  \begin{tabular}{c|c}
    \hline\hline
    Variant & BD-Rate (LD IP32, Class B) \\
    \hline
    No gating                                                   & 18.63\% \\
    Decoder-side spatial gating (no original frame input)       & 17.82\% \\
    Encoder-side spatial gating (gate map transmitted)          & 17.55\% \\
    Scalar gating (ours)                                        & {16.71\%} \\
    \hline\hline
  \end{tabular}
  }
\end{table}

Without any gating, the model always trusts the temporal buffer, yielding $18.63\%$ BD-Rate, the 
same starting point as I3 in Table~\ref{tab:ablation_inter}. Decoder-side spatial gating, which predicts a spatial map without access to the original frame, only improves this to $17.82\%$, because of lacking the information-rich original frame signal. Encoder-side spatial gating, which has access to the original frame but must transmit the predicted map as side information, achieves $17.55\%$. Despite having access to spatial granularity, the encoder-side spatial variant underperforms the scalar gate ($16.71\%$), suggesting that the overhead of spatial gating outweighs its benefits at this scale. Our scalar design strikes the best balance: it conditions on the original frame, incurs negligible bitrate overhead, and achieves the lowest BD-Rate.

\paragraph{Temporal path and training regularization.}
Table~\ref{tab:ablation_temporal} reports additional ablations on the recurrent buffer update, 
classifier supervision, and gate regularization.

\begin{table}[htbp]
  \centering
  \caption{Ablation on temporal path and training regularization on HEVC Class~B, LD IP32. 
  Anchor is VTM 18.0 RA (lower is better).}
  \label{tab:ablation_temporal}
  \begin{tabular}{c|c|c}
    \hline\hline
    Component & w/o & w/ \\
    \hline
    Recurrent buffer update     & 18.77\% & {16.71\%} \\
    Label-smoothed BCE          & 18.77\% & {16.71\%} \\
    Regularization term $\mathcal{L}_\mathrm{reg}$ & 17.21\% & {16.71\%} \\
    \hline\hline
  \end{tabular}
\end{table}

{Recurrent buffer update:} Removing the LSTM-style gated update and replacing 
it with a single-step buffer (i.e., always overwriting the stored state with the current frame 
feature) degrades BD-Rate from $16.71\%$ to $18.77\%$. This confirms that accumulating 
long-range temporal history through selective retention is useful for effective inter 
prediction.

{Label-smoothed BCE supervision:} Removing label-smoothed BCE for classifier training similarly raises BD-Rate to $18.77\%$. Label smoothing BCE 
guides the gate during training and encourages calibrated 
confidence estimates.

{Regularization term $\mathcal{L}_\mathrm{reg}$:} Removing the sparsity-and-smoothness regularizer from loss term increases BD-Rate from $16.71\%$ to 
$17.21\%$. The regularizer discourages gratuitous temporal usage and suppresses erratic frame-to-frame fluctuations in $\alpha_t$, leading to more consistent and robust 
gating behavior across diverse video content.
\section{Conclusion}
In this paper, we propose {Uni-LVC}, a unified learned video codec that supports all-intra, low-delay, and random-access coding in a single model. By formulating inter-coding as \emph{conditional intra} and introducing an efficient hybrid cross-attention module together with a reliability-aware temporal classifier, Uni-LVC avoids the need for mode-specific architectures while remaining robust under unreliable references. Built upon a superior intra backbone with variable-rate capability and trained through a multi-stage curriculum with knowledge replay, Uni-LVC achieves consistent rate–distortion gains and practical latency across coding modes, offering a compact alternative to codecs that require multiple specialized models.

\textbf{Limitations and future work.}
Although effective, Uni-LVC still has several open challenges. First, the temporal reliability gate operates at the frame level; extending it to advanced and efficient spatial or uncertainty-aware modulation may better handle localized motion errors or partial scene changes. Second, while the current design remains efficient at HD resolution, true real-time 4K/8K deployment will require further reductions in attention and memory cost, potentially via pruning~\cite{cheng2024survey} or low-precision inference~\cite{jia2025towards}. Third, the present training and evaluation focus on 8-bit BT.709 RGB; broader support for HDR, higher bit-depths, wider gamuts, and diverse chroma formats is necessary for production use. Another emerging field is optimizing compression for machine-vision tasks. Future work could adapt the Uni-LVC framework to preserve features essential for semantic tasks rather than pixel-level fidelity for human viewing. An exploration of these features, as well as their integration into practical systems, is an exciting direction that we leave for future work.


\bibliographystyle{IEEEtran}
\bibliography{main}

\begin{IEEEbiography}[{\includegraphics[width=1in,height=1.25in,clip,keepaspectratio]{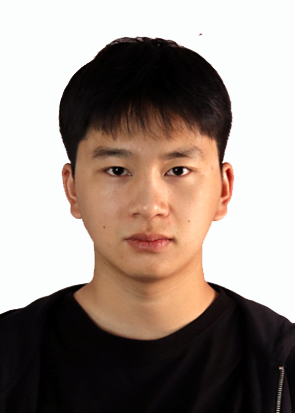}}]{Yichi Zhang}
received a B.E. degree in Computer Science from Hangzhou Normal University, Hangzhou, China, in 2023 and an M.S. degree in Electrical and Computer Engineering from Purdue University, West Lafayette, IN, USA, in 2024. He is now a Ph.D. student at the Video and Image Processing Laboratory at Purdue University, West Lafayette, IN, USA.
His research interests include data compression and neural dynamics.
\end{IEEEbiography}

\begin{IEEEbiography}[{\includegraphics[width=1in,height=1.25in,clip,keepaspectratio]{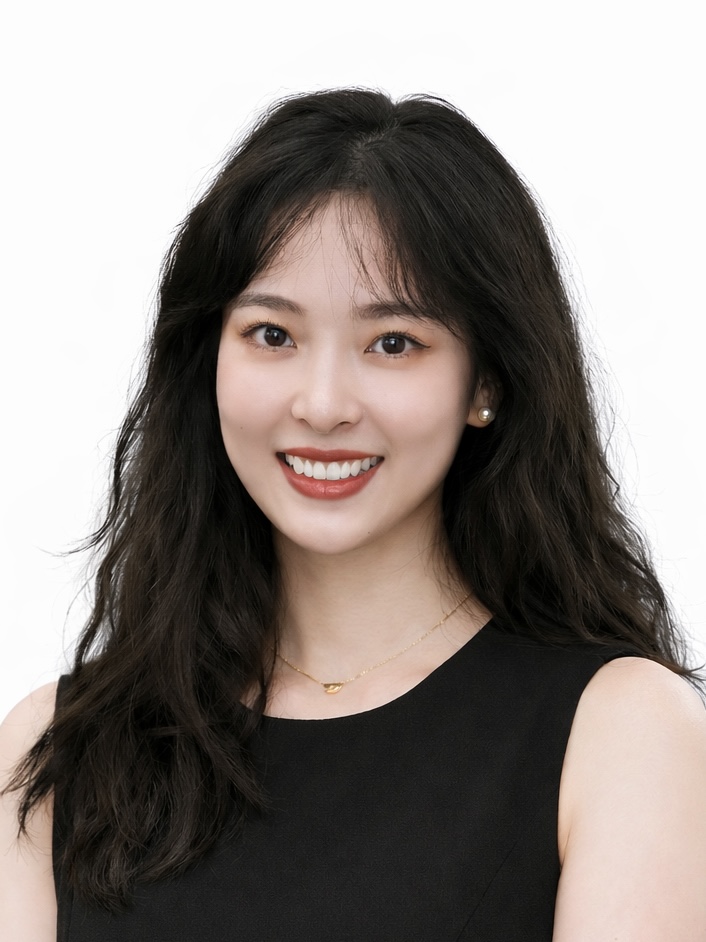}}]{Ruoyu yang} 
received a B.E. degree in Electrical Engineering from Xi'an Jiaotong University, Xi'an, China, in 2023. She is now a Ph.D. student at the Video and Image Processing Laboratory at Purdue University, West Lafayette, IN, USA.
Her research interests include image and video compression.
\end{IEEEbiography}

\begin{IEEEbiography}[{\includegraphics[width=1in,height=1.25in,clip,keepaspectratio]{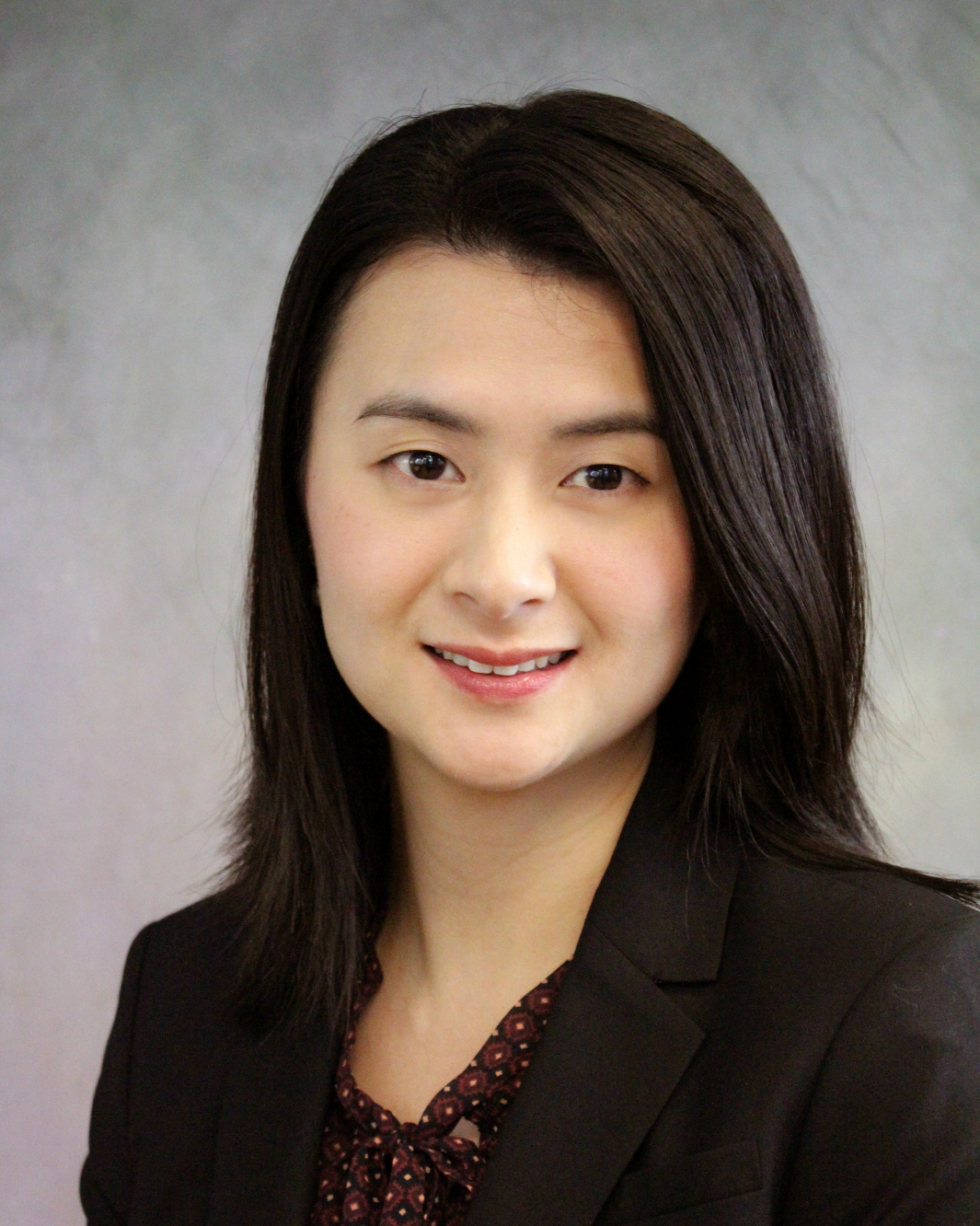}}]{Fengqing Zhu}
(Senior Member, IEEE) received the B.S.E.E. (with highest distinction), M.S., and Ph.D. degrees in Electrical and Computer Engineering from Purdue University in 2004, 2006, and 2011, respectively. From 2012 to 2014, she was a Staff Researcher at Futurewei Technologies, where she received a Certification of Recognition for Core Technology Contribution in 2012. She joined Purdue University, West Lafayette, IN, USA, in 2015, where she is currently an Associate Professor of Electrical and Computer Engineering. Her research interests include visual coding for machines and visual data analytics with a focus on smart health applications. She is a recipient of the 2023 Winter Conference on Applications of Computer Vision Best Algorithms Paper Award and the 2022 Picture Coding Symposium Best Paper Finalist. She is currently serving as the Vice Chair of the IEEE MMSP-TC (2025-2026) and an Elected Member of the IVMSP-TC (2025-2027).
\end{IEEEbiography}

\end{document}